\documentclass[twocolumn,pra,aps,showpacs,superscriptaddress,floatfix,showkeys,nofootinbib,10pt]{revtex4-2}

\usepackage[T1]{fontenc}
\usepackage{amsmath,amsfonts,amssymb,amsthm}
\usepackage{graphicx}
\usepackage{hyperref}

\newcommand{\ket}[1]{ | \, #1 \rangle} \newcommand{\bra}[1]{ \langle #1 \, |}

\newcommand{\be}{\begin{equation}} \newcommand{\ee}{\end{equation}}
\newcommand{\ba}{\begin{aligned}} \newcommand{\ea}{\end{aligned}}

\DeclareMathOperator{\Tr}{Tr}

\DeclareMathOperator{\CHSH}{CHSH}
\DeclareMathOperator{\MCHSH}{MCHSH}
\DeclareMathOperator{\BC3}{BC_3}
\DeclareMathOperator{\I1}{I_1}
\DeclareMathOperator{\h}{h}

\DeclareRobustCommand\openone{\leavevmode\hbox{\small1\normalsize\kern-.33em1}}%

\begin{document}

\title{Bounding conditional entropy of bipartite states with Bell operators}

\author{Jan~Horodecki}
\affiliation{Faculty of Electronics, Telecommunications and Informatics, Gda\'{n}sk University of Technology, 80-233 Gda\'{n}sk, Poland}

\author{Piotr~Mironowicz}
\affiliation{Faculty of Electronics, Telecommunications and Informatics, Gda\'{n}sk University of Technology, 80-233 Gda\'{n}sk, Poland}
\affiliation{International Centre for Theory of Quantum Technologies, University of Gda\'{n}sk, 80-308 Gda\'{n}sk, Poland}
\affiliation{Department of Physics, Stockholm University, S-10691 Stockholm, Sweden}

\date{\today}

\begin{abstract}
Quantum information theory explores numerous properties that surpass classical paradigms, offering novel applications and benefits. Among these properties, negative conditional von~Neumann entropy (CVNE) is particularly significant in entangled quantum systems, serving as an indicator of potential advantages in various information-theoretic tasks, despite its indirect observability. In this paper, we investigate the relationship between CVNE and the violation of Bell inequalities. Our goal is to establish upper bounds on CVNE through semi-definite programming applied to entangled qubits and qutrits, utilizing selected Bell operators.

Our findings reveal that a semi-device-independent certification of negative CVNE is achievable and could be practically beneficial. We further explore two types of robustness: robustness against detection efficiency loopholes, measured by relative violation, and robustness against white noise and imperfections in state preparation, measured by critical visibility. Additionally, we analyze parametrized families of Bell inequalities to identify optimal parameters for different robustness criteria.

This study demonstrates that different Bell inequalities exhibit varying degrees of robustness depending on the desired properties, such as the type of noise resistance or the target level of negative CVNE. By bridging the gap between Bell inequalities and CVNE, our research enhances understanding of the quantum properties of entangled systems and offers insights for practical quantum information processing tasks.
\end{abstract}


\maketitle

\section{Introduction}

Quantum mechanics and quantum information have profoundly expanded our understanding of the fundamental nature of reality, revealing astonishing and often counterintuitive properties. The discovery of Bell inequalities in the 1960s~\cite{bell1964einstein} marked the beginning of quantum information theory, providing a direct example of how quantum resources can surpass classical limits in performing elementary information processing tasks, particularly in generating correlations between subsystems~\cite{horodecki2009quantum}. More recently, these fields have also become powerful tools in cryptography, enabling secure communication and information processing through quantum methods. One particularly intriguing approach in quantum cryptography is device-independent cryptography~\cite{mayers1998quantum}. This approach allows the certification of randomness and the unpredictability of outcomes in cryptographic protocols without relying on detailed knowledge or trust in the underlying devices, instead leveraging the fundamental principles of quantum mechanics to guarantee security. The device-independent framework has been successfully applied in key distribution and randomness generation protocols~\cite{pirandola2020advances}, where Bell inequalities are used as certificates to verify the non-local correlations exhibited by entangled quantum systems.

However, while the device-independent approach has been effective in certifying various aspects of quantum systems, the certification of negative conditional von~Neumann entropy (CVNE) remains largely unexplored. The von~Neumann entropy, a quantum analog of the Shannon entropy, is crucial in quantum information theory and quantum communication. It is used to quantify entanglement between quantum systems and to analyze the security of quantum cryptographic protocols. It also plays a key role in understanding the behavior and capacity of quantum channels for reliable quantum information transmission~\cite{wilde2013quantum}. The discovery that CVNE in entangled quantum systems can take negative values~\cite{cerf1997negative} fundamentally altered our perspective on information flow and the thermodynamics of information processing~\cite{del2011thermodynamic}. Negative conditional information underlies dense coding protocols~\cite{bennett1992communication}, quantum teleportation~\cite{bennett1993teleporting}, and quantum state merging (QSM)~\cite{horodecki2005partial,horodecki2007quantum}. In information theory, conditional entropy measures the uncertainty of one variable given knowledge of another, and negative conditional entropy implies the presence of "negative information," a concept with profound implications.

To date, efficient certification of negative conditional entropy using the device-independent approach has not been fully developed, though preliminary results are available~\cite{arnon2019device}. This gap presents both a challenge and an opportunity for further exploration in quantum information theory. Despite the fundamental significance of these phenomena, there has been no systematic effort to directly connect them. By developing methods to certify negative CVNE and linking it to other quantum non-locality manifestations, we can gain deeper insights into the properties of quantum systems and potentially unlock new cryptographic protocols and applications. This paper investigates the relationship between two key quantum phenomena: Bell inequality violation and negative conditional entropy, which, as explained, are central to the essence of quantum information processing and communication.

Our research operates within the semi-device-independent (SDI) framework~\cite{pawlowski2011semi,pivoluska2021semi}, which explores relationships between directly observable quantities and significant, yet unobservable, quantities, such as the unpredictability of a sequence generated by a process—crucial for cryptography~\cite{dorrendorf2009cryptanalysis}. Unlike the device-independent approach, SDI allows for imposing weak, well-justified constraints on devices, such as limiting the dimension of underlying systems~\cite{brunner2013dimension} or energy consumption~\cite{van2017semi}. Using the dimension constraint assumption, we establish a relationship between the observable quantity (Bell inequality violation) and the unobservable quantity (negative CVNE). This relationship could be foundational for developing new communication and information processing protocols, with potential importance for Information and Communications Technology (ICT). Some consequences of negative CVNE are discussed in~\cite{chatzidimitriou2020experimental,sheshadri2021unconventional}.

The primary tool used in modeling these phenomena is the method of semi-definite programming (SDP)~\cite{sdp,Skrzypczyk2023,mironowicz2024semi,tavakoli2023semidefinite}, particularly the see-saw method~\cite{pal2010maximal}, and relaxation methods that approximate the CVNE function using SDP~\cite{fawzi2018efficient,logapx}. Our results include a direct quantitative relationship indicating how the observed magnitude of Bell inequality violation translates into an upper bound on CVNE between subsystems. Notably, for four selected inequalities, we observed that within a specific range of values, the upper limit for CVNE is negative. This observation means that by detecting the expected value of the Bell operator within a certain range, we can certify that CVNE must be negative.

\section{Methods}

\subsection{Bipartite quantum distributions}

In our scenario, we work with a bipartite quantum state within the tensor product space of two-dimensional complex vector spaces, denoted as $\mathbb{C}^{2}\otimes\mathbb{C}^{2}$. This bipartite state is represented by a density matrix $\rho_{AB}$. The two parties involved, commonly referred to as Alice and Bob, each have access to $m_A$ and $m_B$ different measurements, respectively. These measurements result in binary outcomes. We model these measurements using Positive Operator-Valued Measures (POVMs), denoted as $M^A_{ik}$ for Alice's measurements and $M^B_{jl}$ for Bob's measurements, where $i$ ranges from 1 to $m_A$, $j$ ranges from 1 to $m_B$ (indicating the measurement settings), and $k, l$ can be 0 or 1 (indicating the measurement outcomes).

A specific subset of POVMs is the projective measurements (PMs), characterized by measurement operators that are projectors. Particularly relevant for our work are the binary PMs, which in a two-dimensional Hilbert space, are rank-1 operators. In the Bloch sphere representation, a vector in a two-dimensional space is determined by a pair of parameters, $\theta$ and $\phi$, given by the expression:
\begin{equation}
	\label{eq:Bloch}
	\cos(\theta / 2) \ket{0} + \sin(\theta / 2) e^{i \phi} \ket{1},
\end{equation}
with $\theta \in [0, \pi]$ and $\phi \in [0, 2\pi)$. For binary PMs, the first matrix in the POVM is the projector onto the vector defined in equation~\eqref{eq:Bloch}, and the second matrix is its orthogonal complement.

Both the density matrix $\rho_{AB}$ and the POVM matrices are examples of Hermitian positive semi-definite matrices. In our analysis, we impose these constraints within the SDP framework. The specific constraints are:
\begin{enumerate}
	\item $\rho_{AB} \succeq 0$: The density matrix $\rho_{AB}$ must be positive semi-definite.
	\item $M^A_{ik} \succeq 0$ and $M^B_{jl} \succeq 0$: Similarly, for all $i$ and $k$, the POVM matrices $M^A_{ik}$ and $M^B_{jl}$, corresponding to Alice's and Bob's measurements respectively, must be positive semi-definite.
	\item The completeness conditions $\sum_k M^A_{ik} = \openone$ and $\sum_l M^B_{jl} = \openone$ ensure that the sum of the POVM matrices for each party equals the identity matrix.
	\item The density matrix $\rho_{AB}$ is normalized, meaning its trace is equal to 1, which is necessary for the proper normalization of probabilities.
\end{enumerate}

Using these SDP constraints, we can effectively model and solve various optimization problems in quantum information theory involving bipartite states and their associated measurements. Specifically, the joint probability of Alice and Bob obtaining outcomes $a$ and $b$ for the measurement settings $x$ and $y$, respectively, is calculated as $P(a,b|x,y) = \Tr [\rho_{AB} M^A_{xa} M^B_{yb}]$.

We use the notation $\mathcal{S}_{AB}(d_A, d_B)$ to denote the set of density operators $\sigma_{AB}$ on the composite Hilbert space $\mathcal{H}_A \otimes \mathcal{H}_B$, where $\dim(\mathcal{H}_A) = d_A$ and $\dim(\mathcal{H}_B) = d_B$. Each element $\sigma_{AB} \in \mathcal{S}_{AB}(d_A, d_B)$ satisfies:
\begin{enumerate}
	\item $\sigma_{AB} \succeq 0$ (positivity),
	\item $\Tr (\sigma_{AB}) = 1$ (normalization).
\end{enumerate}

\subsection{Entropy of quantum systems}

The CVNE quantifies the amount of information that remains unknown or missing about one quantum system when the state of another quantum system is known. To comprehend CVNE, it is important to first establish the concept of von~Neumann entropy. Von Neumann's entropy, also referred to as quantum entropy or simply entropy, originates from classical information theory and was extended to the domain of quantum mechanics by John von~Neumann. It serves as a means to quantify the degree of uncertainty or randomness associated with a quantum system. Mathematically, the von~Neumann entropy of a quantum system can be expressed as:
\begin{equation}
	S(\rho) \equiv - \Tr(\rho \log_2(\rho)),
\end{equation}
where, $S(\rho)$ denotes the von~Neumann entropy of the density matrix $\rho$, $\Tr$ represents the trace operation, and $\log_2$ denotes the matrix logarithm with base $2$.

Now, let's consider two quantum systems, $A$ and $B$, described by the joint density matrix $\rho_{AB}$. The CVNE of $\rho_{AB}$ is defined as:
\begin{equation}
	S_{A|B}[\rho_{AB}] \equiv S(\rho_{AB}) - S(\rho_{B}),
\end{equation}
where $S(\rho_{AB})$ represents the von~Neumann entropy of the joint density matrix $\rho_{AB}$ describing both subsystems $A$ and $B$ together, and $S(\rho_B)$ represents the von~Neumann entropy of subsystem $B$ alone. If $S_{A|B}[\rho_{AB}]$ is low, it indicates that knowing the state of subsystem $B$ provides a lot of information about subsystem $A$, and vice versa. For instance, it can be shown that for the two-qubit Werner state $\rho^{wer}_2(p) \equiv \frac{1}{4} \left( \openone_4 - p \sum_{i = 1}^{3} \sigma_i \otimes \sigma_i \right)$ the CVNE $S_{A|B}[\rho^{wer}_2(p)]$ is equal
\begin{equation}
	-3 \frac{1 - p}{4} \log_2{\frac{1 - p}{4}} - \frac{1 + 3 p}{4} \log_2{\frac{1 + 3 p}{4}} - 1,
\end{equation}
and thus the CVNE is negative for $p > 0.747614$~\cite{patro2017non,luc2020quantumness,kumar2023quantum}. Similar formula can be obtained e.g. for Gisin states~\cite{friis2017geometry,luc2020quantumness}.

For any state $\rho_{AB}$ with negative CVNE one may construct a negative CVNE witness as
\begin{equation}
	\label{eq:Wrho}
	W_\rho \equiv -\log \rho_{AB} + \openone_A \otimes \log \rho_B.
\end{equation}
To be more specific, if we take a value $T < 0$ and a bipartite state $\rho_{AB}$ with $S_{A|B}[\rho_{AB}] < 0$ and construct the witness $W_\rho$, then some states $\sigma_{AB}$ with CVNE lower than $T$ could be separated with linear constraint from the state space~\cite[Theorem~3]{witnessing}:
\begin{equation}
	\label{eq:witnessT}
	\Tr [W_\rho \sigma_{AB}] \leq T
	\quad\Rightarrow\quad
	S_{A|B}[\sigma_{AB}] \leq T
	.
\end{equation}
The problem of witnessing the so-called Absolute CVNE Non-Negative class was investigated in~\cite{patro2017non}. Lower and upper bounds on CVNE depending on the entropy were given in~\cite{vempati2022unital}.

\subsection{Bell Operators and self-testing}

In this work, we analyze four specific Bell operators, which quantify the correlations between measurement results in Bell tests. We define these Bell operators based on the outcomes $R_i, R_j \in \{-1, +1\}$ of measurements $M^A_i, M^B_j$ and the correlation
\begin{equation}
	\label{eq:Cxy}
	C_{i,j} = \mathbb{E}(R_i \cdot R_j).
\end{equation}
The first Bell operator we consider is the CHSH operator ($\CHSH$), which is defined as the sum of four correlations:
\begin{equation}
	\CHSH \equiv C_{1,1} + C_{1,2} + C_{2,1} - C_{2,2}.
\end{equation}
This operator originates from the work of Clauser, Horne, Shimony, and Holt~\cite{clauser1969proposed}. For CHSH, the maximal value of this operator allowed in quantum mechanics, i.e. the Tsirelson bound, is $2 \sqrt{2}$, whereas the maximal value possible in classical physics, called the local bound, is $2$. The Tsirelson bound is attained with the maximally entangled state and with the measurements specified by angles $\theta^{A}_1 = 0$ and $\theta^{A}_2 = \pi / 2$ for Alice, and by $\theta^{B}_1 = \pi / 4$ and $\theta^{B}_2 = -\pi / 4$ for Bob; and $\phi = 0$.

The ratio between the Bell value and the local bound and is referred to as the 'violation ratio', or 'relative violation'. This ratio is used to quantify the extent to which a physical system violates the local realism assumption, as described by Bell's inequalities, compared to e.g. the maximum violation predicted by Tsirelson's bound. To be more specific, for the Bell expression value $I$ with the local bound $\beta_C$, the violation ratio is defined as $I / \beta_C$~\cite{masanes2011secure,karczewski2022avenues}. Local correlations are often considered as a free resource, as that can be established between different parts of a quantum system without the need for any shared entanglement~\cite{barrett2005nonlocal,horodecki2013quantumness,vempati2022unital}. Thus the violation ratio is a relevant quantifier of the cost of other resources, such as the negative CVNE considered herein. We note that a drawback of violation ratio is that it is sensitive to rescaling and adding constants to Bell expressions.

An alternative normalized measure for evaluating the required level of violation of a Bell inequality can be considered instead of relative violation. For this purpose, we model the noisy entangled state using the Werner state framework~\cite{werner1989quantum}:
\begin{equation}
	\label{eq:wernerNoise}
	\rho(v) = v \ket{\Phi}\bra{\Phi}_{AB} + (1-v) \frac{\openone_{AB}}{D^2},
\end{equation}
where $v$ represents the visibility, and $\ket{\Phi}_{AB}$ denotes the maximally entangled state of two subsystems, each of dimension $D$. The critical visibility is defined as the minimum value of $v$ required to achieve a specified Bell violation value, assuming perfect implementation of the measurement settings.

Notably, when Bell operators, achieving the Tsirelson bound for maximally entangled states, are expressed as a linear combination of correlators as in Eq.~\eqref{eq:Cxy}, the Bell violation $vT$ is realized if and only if the underlying state has a visibility characterized by $v$. Here, $T$ signifies the Tsirelson bound of the Bell operator. This formulation allows for a precise quantification of the visibility required in the presence of noise to observe quantum correlations that exceed classical bounds. In this work we consider both quantites, violation ratio and visibility, as reference.

If a Bell expression requires only a low relative violation, this suggests that the certification process will be robust against detector inefficiency attacks~\cite{wilms2008local}. This robustness is due to the fact that, with appropriate deterministic outcome assignments~\cite{czechlewski2018influence}, any reduction in the observed Bell violation will not be significant enough to hinder the certification of the negative CVNE. Conversely, if the critical visibility is low, the experimental setup is resilient to white noise and imperfections in state preparation. Depending on the specific context, one can choose the type of robustness that is more pertinent.

Self-testing of a quantum state refers to the process of verifying the properties of a quantum state without having complete knowledge about the measurement devices used~\cite{vsupic2020self}. It allows for the estimation of the state's fidelity and other relevant parameters solely based on the observed measurement outcomes. Self-testing using CHSH allows for the estimation of the state's fidelity without requiring knowledge about the specific measurement devices used by Alice and Bob. It relies on the fact that certain quantum states and measurements are uniquely associated with violating the CHSH inequality. The observation that the maximum violation of the CHSH inequality can only be achieved by utilizing a maximally entangled pair of qubits $\ket{\Psi}_{AB} = \frac{1}{\sqrt{2}} (\ket{00}_{AB} + \ket{11}_{AB})$ was made in~\cite{summers1987maximal,popescu1992states,braunstein1992maximal,tsirelson1993some}. Therefore, by observing the violation of the inequality, one can infer the properties of the underlying quantum state. This observation inspires our approach for robust certification of negative CVNE, as $S_{A|B}[\ket{\Psi}_{AB}] = -1$. It is worth noting that self-testing is not limited to CHSH inequalities, and the self-testing protocols may differ in terms of robustness~\cite{slofstra2011lower,miller2013optimal,ramanathan2018steering,mironowicz2019experimentally,valcarce2020minimum,smania2020experimental,vsupic2020self,valcarce2022self}. There are other Bell inequalities and self-testing protocols that can be used to characterize different aspects of quantum states and measurements. These protocols often involve more complex setups and analyses but provide more detailed information about the quantum system under consideration.

The second Bell operator we examine is denoted as $\BC3$, which is part of the Braunstein-Caves series of inequalities~\cite{braunstein1988information}. The Tsirelson and local bounds are $3 \sqrt{3} \approx 5.1962$ and $4$, respectively. The self-testing properties of this expression were given in~\cite{vsupic2016self}. The $\BC3$ operator is defined by the following expression:
\begin{equation}
	\BC3 \equiv C_{1,1} + C_{1,2} + C_{2,2} + C_{2,3} + C_{3,3} - C_{3,1}.
\end{equation}
The Tsirelson bound for $\BC3$ is attained with the maximally entangled state and with the measurements specified by angles $\theta^{A}_1 = \pi /6$, $\theta^{A}_2 = \pi /2$ and $\theta^{A}_3 = 5 \pi / 6$ for Alice, and by $\theta^{B}_1 = 0$, $\theta^{B}_1 = \pi / 3$ and $\theta^{B}_2 = 2 \pi / 3$ for Bob; and $\phi = 0$.

The third Bell operator we consider is a modified version of the CHSH operator called $\MCHSH$. It includes an additional term and different measurement settings for the party represented by Bob. Its Tsirelson bound is equal to $2 \sqrt{2} + 1$, and the local bound is $3$. Its definition is as follows:
\begin{equation}
	\MCHSH \equiv C_{1,2} + C_{1,3} + C_{2,1} + C_{2,2} - C_{2,3}.
\end{equation}
The modified $\CHSH$ operator $\MCHSH$ was introduced in a previous study~\cite{PRA.88.032319}, where both $\BC3$ and $\MCHSH$ revealed to be very efficient in terms of robust certificated for quantum randomness. They were further analyzed as certificates in~\cite{brown2019framework,xiao2023device}. Its Tsirelson bound is achieved with the maximally entangled state and with the measurements specified by angles $\theta^{A}_1 = 0$ and $\theta^{A}_2 = \pi / 2$ for Alice, and by $\theta^{B}_1 = \pi / 2$, $\theta^{B}_2 = \pi / 4$ and $\theta^{B}_3 = - \pi / 4$ for Bob; and $\phi = 0$.

The fourth Bell operator we analyze is denoted as $\I1$. It is defined by the following expression:
\begin{equation}
	\label{eq:I1}
	\I1 \equiv C_{1,2} - C_{1,3} - C_{2,1} - C_{2,2} + C_{3,1} + C_{3,3} + C_{4,1}.
\end{equation}
The $\I1$ operator was also introduced in the same study~\cite{PRA.88.032319} and further analyzed in~\cite{li2019critical}. Its local bound is $5$, and the Tsirelson bound is $1 + 3 \sqrt{3}$.
The angles defining the optimal measurements are $\theta^{A}_1 = 0$, $\theta^{A}_2 = 4 \pi / 3$, $\theta^{A}_3 = 2 \pi / 3$, and $\theta^{A}_4 = \pi / 2$ for Alice, and by $\theta^{B}_1 = \pi / 2$, $\theta^{B}_2 = \pi / 6$ and $\theta^{B}_3 = 5 \pi / 6$ for Bob; and $\phi = 0$.

These four Bell operators have been revealed to have different robustness against noise for application scenarios of min-entropy certification~\cite{PRA.88.032319} or (non-conditional) von~Neumann entropy certification~\cite{xiao2023device}, and are essential for the analysis in this work.
We examine each of the four Bell operators using three different methods, which are detailed in secs~\ref{sec:meth1}, \ref{sec:meth2}, and~\ref{sec:meth3}.



We also consider one of the families of Bell expressions introduced in~\cite{wooltorton2022tight}. The family is parametrized by $\delta \in (0, \pi/6]$, and is defined with the following formula:
\begin{equation}
	\label{eq:Idelta}
	I_\delta \equiv C(0,0) + \frac{1}{\sin{\delta}} \left( C(0,1) + C(1,0) \right) - \frac{1}{\cos{2 \delta}} C(1,1).
\end{equation}
The Tsirelson bound of $I_\delta$ is
\begin{equation}
	2 \cos^3(\delta) / ( \cos(2\delta) \times \sin(\delta) ),
\end{equation}
whereas the local bound is
\begin{equation}
	-1 + \frac{2}{\sin(\delta)} + \frac{1}{\cos(2\delta)}.
\end{equation}

\subsection{Semi-definite optimization and approximation}

Each of the methods used in this context relies on a mathematical framework of SDP. A common form of SDPs is given by John Watrous~\cite{Watrous11}. For two complex Euclidean spaces $\mathcal{X}$ and $\mathcal{Y}$, a semi-definite program in the Watrous form is defined as a triple $(\Phi, A, B)$, where $\Phi$ is a Hermitian and trace-preserving map from operators on $\mathcal{X}$ to operators on $\mathcal{Y}$, and $A$ and $B$ are Hermitian operators on $\mathcal{X}$ and $\mathcal{Y}$. The so-called primal problem in the Watrous form is:
\begin{align}
	\label{SDP-primal-Watrous}
	\begin{split}
		\text{maximize } &\null \Tr (A X) \\
		\text{subject to } &\null \Phi(X) = B,\\
		&\null X \text{ is positive semi-definite}.
	\end{split}
\end{align}

The values of the Bell operators are multilinear with respect to the state density matrix, Alice's POVMs, and Bob's POVMs. This means that they are linear in each of these three types of terms. To formulate the Bell operators as semi-definite objectives, we employ two different approaches:
\begin{enumerate}
	\item In the \textbf{Fixed measurements} approach, described in secs~\ref{sec:meth1} and~\ref{sec:meth2} for each Bell operator we use the measurements which for the maximally entangled stated reach the Tsirelson bound. Specifically, we use projectors onto computational and Hadamard bases for the $\CHSH$ operator and projectors defined in~\cite{PRA.88.032319} for the other operators. In this approach, the only unknown variable is the density matrix, while the measurements are treated as constants.
	\item The \textbf{Seesaw method} involves cyclic optimization over the three groups of variables: state density matrix, Alice's POVMs, and Bob's POVMs. In each iteration of the optimization, one group of variables is treated as the variable, while the other two groups are treated as constants. This method is described in detail in sec.~\ref{sec:meth3}.
\end{enumerate}
By employing these two approaches, we can effectively formulate the Bell operators as semi-definite objectives, enabling us to analyze their properties and optimize them according to our desired criteria. Comparing the approaches we verify the hypothesis, that for any level of noise, for the states of a given negative CVNE, the optimal measurement maximizing the Bell operator value is equal to those which are optimal also for the noiseless case.

Gauss-Radau quadrature is a numerical integration technique that aims to approximate definite integrals by using a weighted sum of function values at specific points within the integration interval. The parameters of the Gauss-Radau quadrature include the quadrature points, weights, and degree of accuracy. The degree of accuracy in the Gauss-Radau quadrature refers to how well the method approximates the integral. It is determined by the number of quadrature points used in the approximation. The more quadrature points employed, the higher the degree of accuracy achieved. Using the Gauss-Radau quadrature it is possible to express both lower-bounds (denoted as apx${}=-1$) and upper-bounds (denoted as apx${}=+1$) on the matrix logarithm function using SDP~\cite{cvxquad}.

\section{Results}

The generic element of the performed optimizations is the following. The aim is to certify that the value of CVNE is equal to or less than a certain number $H$, $H < 0$.  Suppose that for a given Bell operator $B$ we are able to show that the maximal value it attains over states with CVNE greater or equal $H$ is $\omega_H$. In such case, if we observe $B \geq \omega_H$ then can conclude that the CVNE of the underlying state is equal at most $H$. Thus, if we are able to show this property for $B$, then we obtain a certificate for CVNE.

\subsection{Bell operator value maximization with entropy witnesses bounds and constant measurements}
\label{sec:meth1}

We employ an iterative approach to bound the state space, focusing on one witness $W^{(i)}$ at a time and incorporating the corresponding constraint $\Tr W^{(i)} \sigma_{AB} \geq H$ into the constraint set for subsequent iterations. The iteration process terminates when we can no longer find a state with a conditional entropy lower than $H$. At this point, we consider the maximized value as the maximum operator value for CVNE greater or equal to $H$ with fixed measurements.


To clarify the implications of relation~\eqref{eq:witnessT}, we can draw the following conclusion. Consider any state $\rho_{AB}^{(i)}$ such that $S_{A|B}[\rho_{AB}^{(i)}] < 0$. Let $W^{(i)} = W^{(i)}_{\rho_{AB}^{(i)}}$ be the CVNE witness constructed from $\rho_{AB}^{(i)}$ using~\eqref{eq:Wrho}. According to~\eqref{eq:witnessT}, for any given state $\sigma_{AB}$, if $S_{A|B}[\sigma_{AB}] > H$, then $\Tr [W^{(i)} \sigma_{AB}] > H$. This fact allows us to relax the constraint $S_{A|B}[\sigma_{AB}] > H$.

To proceed, we fix the dimensions $d_A$ and $d_B$ of the subsystems. We choose a target value $H$ that we want to certify as an upper bound on CVNE if a Bell certificate reaches at least a certain value $\omega_H$, which we want to determine. Let $\mathcal{W}^{(i)}$ denote the set of CVNE witnesses used in the $i$-th iteration, where $\mathcal{W}^{(1)} \equiv \emptyset$. In each iteration $i$, we maximize $B$, the Bell violation, as a function of $\sigma_{AB}$ (with measurements fixed to those for the Tsirelson bound) under the following relaxed constraints:
\begin{align}
	\label{eq:relaxation}
	\begin{split}
		\text{maximize } &\null B[\sigma_{AB}] \\
		\text{subject to } &\null \forall W \in \mathcal{W}^{(i)}, \; \Tr(W \sigma_{AB}) \geq H, \\
		\text{over } &\null \sigma_{AB} \in \mathcal{S}_{AB}(d_A, d_B).
	\end{split}
\end{align}

Let $\omega^{(i)}$ be the value obtained from this maximization problem~\eqref{eq:relaxation}, and let $\sigma_{AB}^{(i)} \in \mathcal{S}_{AB}(d_A, d_B)$ be the state that achieves the Bell value $\omega^{(i)}$. The relaxed constraints are satisfied by all states $\sigma_{AB}^{(i)} \in \mathcal{S}_{AB}(d_A, d_B)$ such that $S_{A|B}[\sigma_{AB}^{(i)}] \geq H$, and potentially by some states with a CVNE less than $H$. Thus, $\omega^{(i)} \geq \omega_H$. As $\mathcal{W}^{(i)} \subseteq \mathcal{W}^{(i+1)}$, the optimization region becomes more restrictive with increasing $i$, leading to a decrease in the values of $\omega^{(i)}$, i.e., $\omega^{(i+1)} \leq \omega^{(i)}$. If the iterative process is stopped at some $\tilde{i}$, then $\omega_H \leq \omega^{(\tilde{i})}$, and the observation of the Bell value $\omega^{(\tilde{i})}$ ensures that the CVNE is at most $H$.

If, for some $i$, it turns out that $S_{A|B}[\sigma_{AB}^{(i)}] = H$, we conclude that $\omega_H = \omega^{(i)}$ and terminate the iterative process. Otherwise, we set $\rho_{AB}^{(i)} = \sigma_{AB}^{(i)}$ and update $\mathcal{W}^{(i + 1)} \equiv \mathcal{W}^{(i)} \cup \{ W^{(i)} \}$, with $W^{(i)}$ calculated from $\sigma_{AB}^{(i)}$ with~\eqref{eq:Wrho}, continuing the iterative process until the criterion $S_{A|B}[\sigma_{AB}^{(i)}] = H$ or another stopping condition is met.

In cases where it is not possible to construct a witness for a given state $\rho_{AB}^{(i)}$ due to eigenvalues approaching zero, (which is common for pure entangled states), we resort to bisection over states $c \cdot \rho_{AB} + (1-c) \frac{\openone_{AB}}{4}$, where $c \in [0,1]$. This enables us to find a mixed state with slightly lower entropy than $H$ and construct a witness and constraint for that state instead. The existence of such $c$ helps us circumvent numerical issues within the range $[H-10^{-9}, H)$.

\subsection{Operator value maximization with entropy approximations and constant measurements}
\label{sec:meth2}

The second method does not involve iteratively building up constraints. Instead, it utilizes a semi-definite approximation of CVNE to bound the state space in an SDP with fixed measurements. This technique is based on the SDP approximation of CVNE~\cite{logapx,fawzi2018efficient}. In our framework, we incorporate the cvxquad library~\cite{cvxquad}, which leverages the Gauss-Radau quadrature to achieve two possible approximations: a lower bound and an upper bound on CVNE. For our experiments, we set both the $m$ and $k$ parameters of the approximation to $3$. We execute each experiment using this semi-definite approximation twice: once for the lower bound (apx${}=-1$) and once for the upper bound (apx${}=+1$) approximation. The results exhibit only slight differences, indicating that the contribution of the approximation to the overall error is minimal.

\subsection{Operator value maximization with entropy approximations -- see-saw}
\label{sec:meth3}

The third method differs from the previous two in that it is a heuristic approach, which means it may converge to local maxima. We categorize the variables into three groups: the state, Alice's POVMs, and Bob's POVMs. We begin with the state as the variable and treat the remaining groups as constants and initialize them randomly. Without loss of generality, we can assume that the computational basis for both Alice and Bob corresponds to the eigenbasis of their first measurement operators. Furthermore, we restrict these measurements to be projective since, in the case of binary outcomes with qubits, projectors are sufficient to achieve the maximum violation of a Bell operator \cite{vertesi2010two}. This assumption helps improve the numerical convergence of the see-saw computations.

The see-saw optimization proceeds by cycling through steps until the absolute difference between the maximized values in two consecutive cycles is smaller than $10^{-7}$. Similar to the previous method, we utilize the SDP approximations of CVNE to bound it to the given value $H$ in each sub-optimization of the see-saw algorithm. We repeat each see-saw optimization procedure $100$ times, each time initializing the measurements with new random values. We perform this process twice, once for the lower bound (apx${}=-1$) approximation and once for the upper bound (apx${}=+1$) approximation, resulting in a total of $200$ iterations for each $H$ value. The initial randomization of the measurements is performed as follows: four real numbers are drawn from a uniform distribution on the interval $[0,1)$ to form a vector $v \in \mathbb{C}^2$. From this vector, we construct the projectors $P_1 = vv^\dagger$ and $P_2 = \openone - P_1$. For each value of $H$ and apx${}=\pm1$, we repeat the see-saw optimization procedure $100$ times and select the maximum attained Bell operator value as the result.

\subsection{Disussion of the results}

\begin{figure*}[htbp]
	\includegraphics{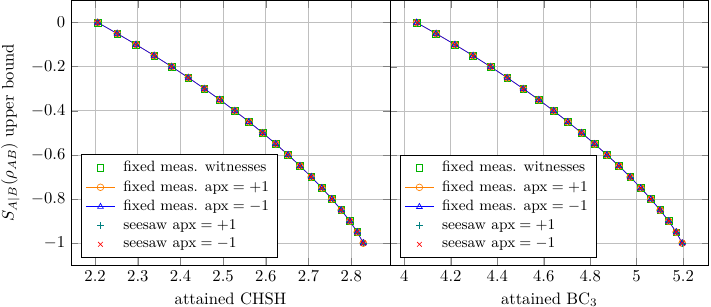}
	\includegraphics{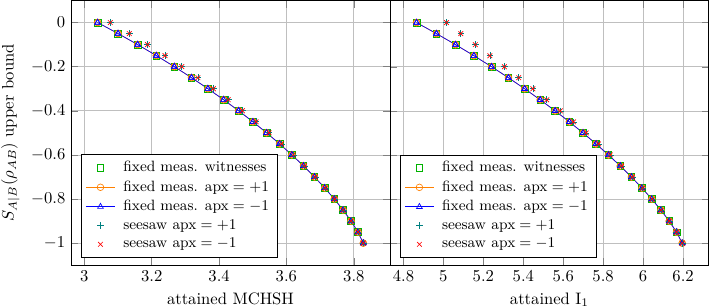}
	\caption{(Color online) Conditional entropy upper bounds for two qubits bipartite state given that we measured some $x$ as Bell operator value. For methods with fixed measurement, the results coincide. A substantial difference emerges between fixed measurements and arbitrary measurements (see-saw) for $\MCHSH$ and $\I1$ operators.}\label{fig:plots}
\end{figure*}

Fig.~\ref{fig:plots} illustrates the relationship between the entropy upper bound and the achieved values of the Bell operator. The obtained results remain consistent for both fixed and arbitrary measurements in the case of $\CHSH$. However, slight differences are observed for $\BC3$ when the CVNE value is close to $0$. Notably, more significant deviations are noticeable for $\MCHSH$ and $\I1$, indicating that the measurements optimal for maximum violation are not necessarily optimal for states with constrained CVNE. This means that the bounds assuming a particular form of the measurements are stronger than the bounds without this presumption, suggesting the possibility of improvement in the form of the source-device-independent certification~\cite{marangon2017source,avesani2018source}. It is worth noting that the see-saw method proved to be considerably slower for operators with a larger number of measurement settings, particularly for $\I1$. In such cases, a single see-saw trial often required several thousand optimization cycles to converge. 

The role of CVNE was elucidated in~\cite{horodecki2005partial}, where the problem of transferring an unknown quantum state distributed over two quantum systems to a single system using quantum communication was investigated. The amount of quantum communication needed for this task is quantified with CVNE, which had been previously understood theoretically but lacked a practical interpretation. If CVNE is positive, the sender needs to communicate this amount of quantum bits to the receiver. However, when CVNE is negative, both the sender and receiver gain a corresponding potential for future quantum communication. To achieve an optimal transfer of partial information, the so-called QSM protocol, which efficiently transfers partial information given by CVNE from one system to another, can be used. The results provided in Fig.~\ref{fig:plots} show that for separated parties sharing an entangled state of a given upper bound on the conditional entropy, the QSM protocol can be applied. Thus the upper bounds based on Bell expression values provide an SDI certificate on the performance of the QSM protocol, or super dense coding protocols~\cite{bennett1992communication,bruss2004distributed,prabhu2013exclusion}.

\begin{table}[htbp]
	\begin{tabular}{|l|l|l|l|l|}
		\hline
		\shortstack{Bell \\ operator} & \shortstack{local \\ bound} & \shortstack{viol. ratio for \\ negative CVNE} & \shortstack{viol. ratio for \\ -0.9 CVNE} & \shortstack{critical \\ visiblity}  \\ \hline
		$\CHSH$  & $2$ & $1.1030$ ($2.2060$) & $1.3984$ ($2.7967$) & $0.9888$  \\ \hline
		$\MCHSH$ & $3$ & $1.0258$ ($3.0773$) & $1.2641$ ($3.7923$) & $0.9906$  \\ \hline
		$\BC3$   & $4$ & $1.0140$ ($4.0559$) & $1.0276$ ($5.1379$) & $0.9888$  \\ \hline
		$I1$     & $5$ & $1.0031$ ($5.0155$) & $1.0219$ ($6.1315$) & $0.9896$  \\ \hline
	\end{tabular}
	\caption{The violation ratio for the four investigated Bell operators needed to obtain a certification that the conditional von~Neumann entropy (CVNE) is negative (column~III), or less than $-0.9$ (columns~IV). The values in the parentheses show the values of Bell operators needed for the relevant cases. The last columns shows the critical visiblity for $-0.9$ CVNE certification, as a reference. It can be observed that the robustness in terms of relative violation of the other Bell operators has been significantly improved compared to $\CHSH$, but on the other hand, the critical visiblity for all Bell operators remains almost the same.}
	\label{tab:BellRatio}
\end{table}

The results for the case of CHSH comply with those stated in~\cite[Lemma~5]{arnon2019device} where the upper bound on CVNE of the form
\begin{equation}
	\label{eq:boundArnon}
	2 \h \left( \frac{1}{2} - \frac{\sqrt{2}}{8} \cdot I \right) - 1,
\end{equation}
where $\h(\cdot)$ is the binary entropy, was obtained using different methods, cf.~\cite[eq.~(13)]{pironio2009device}. Our calculations show that the bound~\eqref{eq:boundArnon} is not only tight but also exact (up to numerical precision). We tried curve fitting analysis to reproduce the numerical results of Fig.~\ref{fig:plots} for the function of the form $p_1 \cdot \h(p_2 - p_3 \cdot I) - p_4$, but the algorithm did not converge. In Tab.~\ref{tab:BellRatio} we compare the violation ratio required to certify the two cases: when the CVNE is negative, and when it is less than $-0.9$. It can be observed that for the three Bell operators $\MCHSH$, $\BC3$, and $\I1$, the violation ratio is in both cases significantly lower than for $\CHSH$, but the critical visibility for all Bell operators remain very similar. Recent experiments indicate that the visiblitites above $0.99$, as needed for certification of $-0.9$ bits of CVNE, are feasible with the current technology~\cite{seguinard2023experimental}. This shows that the robustness of the certification of negative CVNE may be facilitated, e.g. using a proper post-processing~\cite{czechlewski2018influence}. This robustness for SDI certification plays the role of one-shot distillable entanglement protocols~\cite{arnon2019device} or distributed private randomness distillation~\cite{yang2019distributed}.

\begin{figure}[htbp]
	\centering
	\includegraphics[width=0.97\linewidth]{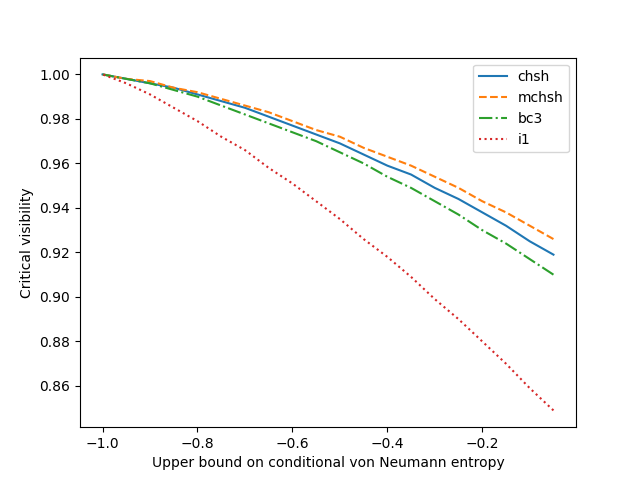}
	\caption{(Color online) Robustness of the certification of negative conditional von~Neumann entropy using four different Bell inequalities, shown as a function of visibility. The upper bound on the dimension of the entangled state is set at two qutrits. The most robust inequality in terms of visibility is $I_1$, as defined in Eq.~\eqref{eq:I1}.}
	\label{fig:dim3x3plot}
\end{figure}

We conducted similar optimizations for a state entangling two qutrits. Figure~\ref{fig:dim3x3plot} illustrates the robustness of the certification of negative CVNE using four different Bell inequalities, plotted as a function of visibility. The most robust inequality, in terms of visibility, is $I_1$, as defined in~\eqref{eq:I1}. Comparing these results with Table~\ref{tab:BellRatio}, we observe that as the dimension of the entangled state increases, different Bell certificates may prove to be the most robust.

Next, we address the problem of identifying the Bell operator best suited for our scenario, either in terms of relative Bell violation or critical visibility, depending on the desired certification of negative CVNE. Specifically, we examine the family of Bell expressions given in~\eqref{eq:Idelta}. We determine the optimal value of the parameter $\delta$ based on the desired robustness properties.

\begin{figure}[htbp]
	\centering
	\includegraphics[width=0.97\linewidth]{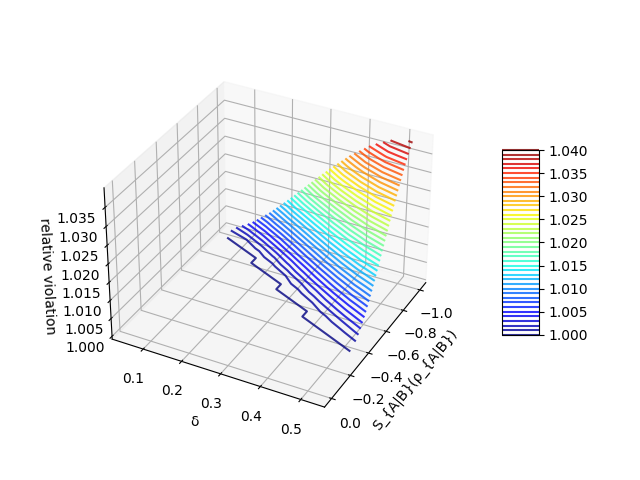}
	\caption{(Color online) Required relative violation of various Bell inequalities from the family~\eqref{eq:Idelta} to certify the negativity of conditional von~Neumann entropy. Lower relative violation indicates greater robustness against the detection efficiency loophole, making the protocol more resilient. Smaller values of $\delta$ correspond to higher robustness in these terms.}
	\label{fig:relativeviolationidelta3d}
\end{figure}

In Figure~\ref{fig:relativeviolationidelta3d}, the relative violation of various Bell inequalities from the family~\eqref{eq:Idelta} is presented, which is necessary to certify the negativity of CVNE. A smaller required relative violation indicates greater ease in circumventing the detection efficiency loophole, thereby making the protocol more robust. Hence, smaller values of $\delta$ exhibit greater robustness in this regard.

\begin{figure}[htbp]
	\centering
	\includegraphics[width=0.97\linewidth]{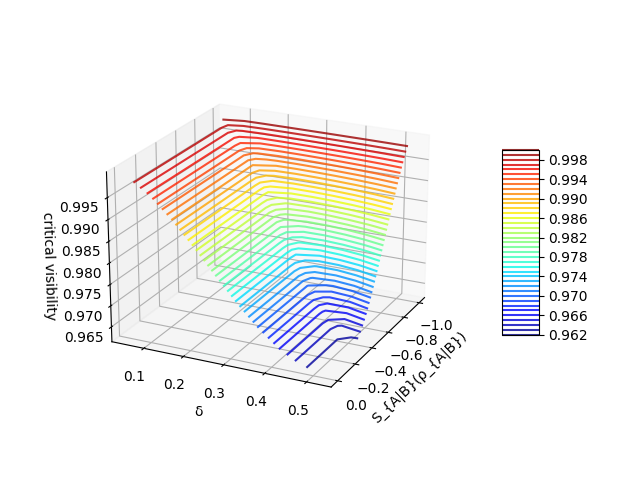}
	\caption{(Color online) Critical visibility required to certify the negativity of conditional von~Neumann entropy using the Bell expressions from the family~\eqref{eq:Idelta}. Lower critical visibility indicates greater robustness of the certification against white noise and imperfections in the preparation of the entangled state. The results show that larger values of $\delta$ provide more robust certification.}
	\label{fig:criticalvisivilityidelta3d}
\end{figure}

Figure~\ref{fig:criticalvisivilityidelta3d} shows the critical visibility required to certify negative CVNE using the Bell expressions from the same family~\eqref{eq:Idelta}. Lower critical visibility implies that the certification is more robust against white noise and imperfections in the preparation of the entangled state. It can be observed that larger values of $\delta$ provide greater robustness. Therefore, by comparing Figures~\ref{fig:relativeviolationidelta3d} and \ref{fig:criticalvisivilityidelta3d}, we conclude that depending on the type of robustness sought and the desired value of certified negative CVNE, different representatives from the family~\eqref{eq:Idelta} may be optimal.
In particular, from Figure~\ref{fig:relativeviolationidelta3d} we conclude that the $I_\delta$ certificate is most robust in terms of required relative violation for low values of $\delta$ parameter. On the other hand, from Figure~\ref{fig:criticalvisivilityidelta3d} we see that the robustness in terms of required critical visibility is optimal for high values of $\delta$ parameter.

\section{Conclusions}

In this paper, we presented an analysis of the relationship between the violation of four selected Bell inequalities and the upper bound on the conditional entropy of the corresponding quantum states. To achieve this, we maximized each Bell operator over all possible entangled two-part states, specifically considering both qubits and qutrits, with the condition that the conditional von~Neumann entropy (CVNE) is greater than a specific threshold. Our findings demonstrate that if a Bell value exceeds a determined threshold, CVNE must necessarily be less than the specified value. Consequently, for each investigated inequality, there exists a range of values that certify the necessity of negative CVNE.

We explored two types of robustness across four different Bell inequalities: robustness against detection efficiency attacks (measured by relative violation) and robustness against white noise and state preparation imperfections (measured by critical visibility). Additionally, we examined the behavior of parametrized families of Bell inequalities in terms of these robustness measures. This dual examination provides a nuanced understanding of how different Bell inequalities respond to different types of noise and imperfections in experimental setups.

Our analysis shows that the violation of a Bell inequality indicates the presence of entanglement in the quantum system. Although all states with negative CVNE are entangled~\cite{cerf1999quantum}, it is crucial to note that not all entangled states exhibit negative CVNE. Therefore, our investigation offers a quantitative distinction and interconnection between these two quantum phenomena. This approach complements the work of~\cite{friis2017geometry,friis2019entanglement}, where the conditional amplitude operator~\cite{cerf1997quantum} was used to infer the existence of entanglement from the negativity of CVNE, akin to the PPT criterion~\cite{HORODECKI19961,peres1996separability}.

As anticipated, certification of negative CVNE was only possible above the classical threshold for violating the inequalities. Moreover, since each considered inequality possesses the self-test property for the maximally entangled state~\cite{vsupic2020self}, the certified negativity reached a value of $-1$ at maximum violation. Interestingly, the threshold for certifying negativity was consistently above the classical bound, indicating that the relationship between Bell violation and negative CVNE is non-trivial. However, the different Bell operators exhibited varying degrees of robustness depending on the type of robustness considered.

Our study assumed that the entangled quantum state has a dimension of two or three, situating our work within a semi-device-independent framework. Future research could systematically explore the analogous relationship between the value of the Bell operator and the required negativity of conditional entropy to achieve full device-independent certification. Additionally, it would be valuable to identify Bell inequalities that are more robust for certification purposes and to investigate whether any Bell inequalities can be maximally violated without resulting in negative CVNE.

\section*{Acknowledgements}

This work was supported by the Knut and Alice Wallenberg Foundation through the Wallenberg Centre for Quantum Technology (WACQT), and NCBiR QUANTERA/2/2020 (www.quantera.eu) an ERA-Net co-fund in Quantum Technologies under the project eDICT. Certain computations were conducted on a server financed by 2018/MAB/5/AS-1.

The source code used for conducting the experiments in this work is publicly available to be accessed and downloaded from the following URL: \url{https://git.pg.edu.pl/eti/bellop_cond_entropy}. It has been implemented using libraries~\cite{picos, cvxopt, mosek}.


\begin{thebibliography}{77}%
\makeatletter
\providecommand \@ifxundefined [1]{%
 \@ifx{#1\undefined}
}%
\providecommand \@ifnum [1]{%
 \ifnum #1\expandafter \@firstoftwo
 \else \expandafter \@secondoftwo
 \fi
}%
\providecommand \@ifx [1]{%
 \ifx #1\expandafter \@firstoftwo
 \else \expandafter \@secondoftwo
 \fi
}%
\providecommand \natexlab [1]{#1}%
\providecommand \enquote  [1]{``#1''}%
\providecommand \bibnamefont  [1]{#1}%
\providecommand \bibfnamefont [1]{#1}%
\providecommand \citenamefont [1]{#1}%
\providecommand \href@noop [0]{\@secondoftwo}%
\providecommand \href [0]{\begingroup \@sanitize@url \@href}%
\providecommand \@href[1]{\@@startlink{#1}\@@href}%
\providecommand \@@href[1]{\endgroup#1\@@endlink}%
\providecommand \@sanitize@url [0]{\catcode `\\12\catcode `\$12\catcode
  `\&12\catcode `\#12\catcode `\^12\catcode `\_12\catcode `\%12\relax}%
\providecommand \@@startlink[1]{}%
\providecommand \@@endlink[0]{}%
\providecommand \url  [0]{\begingroup\@sanitize@url \@url }%
\providecommand \@url [1]{\endgroup\@href {#1}{\urlprefix }}%
\providecommand \urlprefix  [0]{URL }%
\providecommand \Eprint [0]{\href }%
\providecommand \doibase [0]{https://doi.org/}%
\providecommand \selectlanguage [0]{\@gobble}%
\providecommand \bibinfo  [0]{\@secondoftwo}%
\providecommand \bibfield  [0]{\@secondoftwo}%
\providecommand \translation [1]{[#1]}%
\providecommand \BibitemOpen [0]{}%
\providecommand \bibitemStop [0]{}%
\providecommand \bibitemNoStop [0]{.\EOS\space}%
\providecommand \EOS [0]{\spacefactor3000\relax}%
\providecommand \BibitemShut  [1]{\csname bibitem#1\endcsname}%
\let\auto@bib@innerbib\@empty
\bibitem [{\citenamefont {Bell}(1964)}]{bell1964einstein}%
  \BibitemOpen
  \bibfield  {author} {\bibinfo {author} {\bibfnamefont {J.~S.}\ \bibnamefont
  {Bell}},\ }\bibfield  {title} {\bibinfo {title} {On the {E}instein {P}odolsky
  {R}osen paradox},\ }\href@noop {} {\bibfield  {journal} {\bibinfo  {journal}
  {Physics Physique Fizika}\ }\textbf {\bibinfo {volume} {1}},\ \bibinfo
  {pages} {195} (\bibinfo {year} {1964})}\BibitemShut {NoStop}%
\bibitem [{\citenamefont {Horodecki}\ \emph {et~al.}(2009)\citenamefont
  {Horodecki}, \citenamefont {Horodecki}, \citenamefont {Horodecki},\ and\
  \citenamefont {Horodecki}}]{horodecki2009quantum}%
  \BibitemOpen
  \bibfield  {author} {\bibinfo {author} {\bibfnamefont {R.}~\bibnamefont
  {Horodecki}}, \bibinfo {author} {\bibfnamefont {P.}~\bibnamefont
  {Horodecki}}, \bibinfo {author} {\bibfnamefont {M.}~\bibnamefont
  {Horodecki}},\ and\ \bibinfo {author} {\bibfnamefont {K.}~\bibnamefont
  {Horodecki}},\ }\bibfield  {title} {\bibinfo {title} {Quantum entanglement},\
  }\href@noop {} {\bibfield  {journal} {\bibinfo  {journal} {Reviews of modern
  physics}\ }\textbf {\bibinfo {volume} {81}},\ \bibinfo {pages} {865}
  (\bibinfo {year} {2009})}\BibitemShut {NoStop}%
\bibitem [{\citenamefont {Mayers}\ and\ \citenamefont
  {Yao}(1998)}]{mayers1998quantum}%
  \BibitemOpen
  \bibfield  {author} {\bibinfo {author} {\bibfnamefont {D.}~\bibnamefont
  {Mayers}}\ and\ \bibinfo {author} {\bibfnamefont {A.}~\bibnamefont {Yao}},\
  }\bibfield  {title} {\bibinfo {title} {Quantum cryptography with imperfect
  apparatus},\ }in\ \href@noop {} {\emph {\bibinfo {booktitle} {Proceedings
  39th Annual Symposium on Foundations of Computer Science (Cat. No.
  98CB36280)}}}\ (\bibinfo {organization} {IEEE},\ \bibinfo {year} {1998})\
  pp.\ \bibinfo {pages} {503--509}\BibitemShut {NoStop}%
\bibitem [{\citenamefont {Pirandola}\ \emph {et~al.}(2020)\citenamefont
  {Pirandola}, \citenamefont {Andersen}, \citenamefont {Banchi}, \citenamefont
  {Berta}, \citenamefont {Bunandar}, \citenamefont {Colbeck}, \citenamefont
  {Englund}, \citenamefont {Gehring}, \citenamefont {Lupo}, \citenamefont
  {Ottaviani}, \citenamefont {Pereira}, \citenamefont {Razavi}, \citenamefont
  {Shamsul~Shaari}, \citenamefont {Tomamichel}, \citenamefont {Usenko},
  \citenamefont {Vallone}, \citenamefont {Villoresi},\ and\ \citenamefont
  {Wallden}}]{pirandola2020advances}%
  \BibitemOpen
  \bibfield  {author} {\bibinfo {author} {\bibfnamefont {S.}~\bibnamefont
  {Pirandola}}, \bibinfo {author} {\bibfnamefont {U.~L.}\ \bibnamefont
  {Andersen}}, \bibinfo {author} {\bibfnamefont {L.}~\bibnamefont {Banchi}},
  \bibinfo {author} {\bibfnamefont {M.}~\bibnamefont {Berta}}, \bibinfo
  {author} {\bibfnamefont {D.}~\bibnamefont {Bunandar}}, \bibinfo {author}
  {\bibfnamefont {R.}~\bibnamefont {Colbeck}}, \bibinfo {author} {\bibfnamefont
  {D.}~\bibnamefont {Englund}}, \bibinfo {author} {\bibfnamefont
  {T.}~\bibnamefont {Gehring}}, \bibinfo {author} {\bibfnamefont
  {C.}~\bibnamefont {Lupo}}, \bibinfo {author} {\bibfnamefont {C.}~\bibnamefont
  {Ottaviani}}, \bibinfo {author} {\bibfnamefont {J.~L.}\ \bibnamefont
  {Pereira}}, \bibinfo {author} {\bibfnamefont {M.}~\bibnamefont {Razavi}},
  \bibinfo {author} {\bibfnamefont {J.}~\bibnamefont {Shamsul~Shaari}},
  \bibinfo {author} {\bibfnamefont {M.}~\bibnamefont {Tomamichel}}, \bibinfo
  {author} {\bibfnamefont {V.~C.}\ \bibnamefont {Usenko}}, \bibinfo {author}
  {\bibfnamefont {G.}~\bibnamefont {Vallone}}, \bibinfo {author} {\bibfnamefont
  {P.}~\bibnamefont {Villoresi}},\ and\ \bibinfo {author} {\bibfnamefont
  {P.}~\bibnamefont {Wallden}},\ }\bibfield  {title} {\bibinfo {title}
  {Advances in quantum cryptography},\ }\href@noop {} {\bibfield  {journal}
  {\bibinfo  {journal} {Advances in optics and photonics}\ }\textbf {\bibinfo
  {volume} {12}},\ \bibinfo {pages} {1012} (\bibinfo {year}
  {2020})}\BibitemShut {NoStop}%
\bibitem [{\citenamefont {Wilde}(2017)}]{wilde2013quantum}%
  \BibitemOpen
  \bibfield  {author} {\bibinfo {author} {\bibfnamefont {M.~M.}\ \bibnamefont
  {Wilde}},\ }\href {https://doi.org/10.1017/9781316809976} {\emph {\bibinfo
  {title} {Quantum information theory}}},\ \bibinfo {edition} {2nd}\ ed.\
  (\bibinfo  {publisher} {Cambridge University Press},\ \bibinfo {year}
  {2017})\BibitemShut {NoStop}%
\bibitem [{\citenamefont {Cerf}\ and\ \citenamefont
  {Adami}(1997)}]{cerf1997negative}%
  \BibitemOpen
  \bibfield  {author} {\bibinfo {author} {\bibfnamefont {N.~J.}\ \bibnamefont
  {Cerf}}\ and\ \bibinfo {author} {\bibfnamefont {C.}~\bibnamefont {Adami}},\
  }\bibfield  {title} {\bibinfo {title} {Negative entropy and information in
  quantum mechanics},\ }\href@noop {} {\bibfield  {journal} {\bibinfo
  {journal} {Physical Review Letters}\ }\textbf {\bibinfo {volume} {79}},\
  \bibinfo {pages} {5194} (\bibinfo {year} {1997})}\BibitemShut {NoStop}%
\bibitem [{\citenamefont {Del~Rio}\ \emph {et~al.}(2011)\citenamefont
  {Del~Rio}, \citenamefont {{\AA}berg}, \citenamefont {Renner}, \citenamefont
  {Dahlsten},\ and\ \citenamefont {Vedral}}]{del2011thermodynamic}%
  \BibitemOpen
  \bibfield  {author} {\bibinfo {author} {\bibfnamefont {L.}~\bibnamefont
  {Del~Rio}}, \bibinfo {author} {\bibfnamefont {J.}~\bibnamefont {{\AA}berg}},
  \bibinfo {author} {\bibfnamefont {R.}~\bibnamefont {Renner}}, \bibinfo
  {author} {\bibfnamefont {O.}~\bibnamefont {Dahlsten}},\ and\ \bibinfo
  {author} {\bibfnamefont {V.}~\bibnamefont {Vedral}},\ }\bibfield  {title}
  {\bibinfo {title} {The thermodynamic meaning of negative entropy},\
  }\href@noop {} {\bibfield  {journal} {\bibinfo  {journal} {Nature}\ }\textbf
  {\bibinfo {volume} {474}},\ \bibinfo {pages} {61} (\bibinfo {year}
  {2011})}\BibitemShut {NoStop}%
\bibitem [{\citenamefont {Bennett}\ and\ \citenamefont
  {Wiesner}(1992)}]{bennett1992communication}%
  \BibitemOpen
  \bibfield  {author} {\bibinfo {author} {\bibfnamefont {C.~H.}\ \bibnamefont
  {Bennett}}\ and\ \bibinfo {author} {\bibfnamefont {S.~J.}\ \bibnamefont
  {Wiesner}},\ }\bibfield  {title} {\bibinfo {title} {Communication via one-and
  two-particle operators on {E}instein-{P}odolsky-{R}osen states},\ }\href@noop
  {} {\bibfield  {journal} {\bibinfo  {journal} {Physical Review Letters}\
  }\textbf {\bibinfo {volume} {69}},\ \bibinfo {pages} {2881} (\bibinfo {year}
  {1992})}\BibitemShut {NoStop}%
\bibitem [{\citenamefont {Bennett}\ \emph {et~al.}(1993)\citenamefont
  {Bennett}, \citenamefont {Brassard}, \citenamefont {Cr{\'e}peau},
  \citenamefont {Jozsa}, \citenamefont {Peres},\ and\ \citenamefont
  {Wootters}}]{bennett1993teleporting}%
  \BibitemOpen
  \bibfield  {author} {\bibinfo {author} {\bibfnamefont {C.~H.}\ \bibnamefont
  {Bennett}}, \bibinfo {author} {\bibfnamefont {G.}~\bibnamefont {Brassard}},
  \bibinfo {author} {\bibfnamefont {C.}~\bibnamefont {Cr{\'e}peau}}, \bibinfo
  {author} {\bibfnamefont {R.}~\bibnamefont {Jozsa}}, \bibinfo {author}
  {\bibfnamefont {A.}~\bibnamefont {Peres}},\ and\ \bibinfo {author}
  {\bibfnamefont {W.~K.}\ \bibnamefont {Wootters}},\ }\bibfield  {title}
  {\bibinfo {title} {Teleporting an unknown quantum state via dual classical
  and {E}instein-{P}odolsky-{R}osen channels},\ }\href@noop {} {\bibfield
  {journal} {\bibinfo  {journal} {Physical Review Letters}\ }\textbf {\bibinfo
  {volume} {70}},\ \bibinfo {pages} {1895} (\bibinfo {year}
  {1993})}\BibitemShut {NoStop}%
\bibitem [{\citenamefont {Horodecki}\ \emph {et~al.}(2005)\citenamefont
  {Horodecki}, \citenamefont {Oppenheim},\ and\ \citenamefont
  {Winter}}]{horodecki2005partial}%
  \BibitemOpen
  \bibfield  {author} {\bibinfo {author} {\bibfnamefont {M.}~\bibnamefont
  {Horodecki}}, \bibinfo {author} {\bibfnamefont {J.}~\bibnamefont
  {Oppenheim}},\ and\ \bibinfo {author} {\bibfnamefont {A.}~\bibnamefont
  {Winter}},\ }\bibfield  {title} {\bibinfo {title} {Partial quantum
  information},\ }\href@noop {} {\bibfield  {journal} {\bibinfo  {journal}
  {Nature}\ }\textbf {\bibinfo {volume} {436}},\ \bibinfo {pages} {673}
  (\bibinfo {year} {2005})}\BibitemShut {NoStop}%
\bibitem [{\citenamefont {Horodecki}\ \emph {et~al.}(2007)\citenamefont
  {Horodecki}, \citenamefont {Oppenheim},\ and\ \citenamefont
  {Winter}}]{horodecki2007quantum}%
  \BibitemOpen
  \bibfield  {author} {\bibinfo {author} {\bibfnamefont {M.}~\bibnamefont
  {Horodecki}}, \bibinfo {author} {\bibfnamefont {J.}~\bibnamefont
  {Oppenheim}},\ and\ \bibinfo {author} {\bibfnamefont {A.}~\bibnamefont
  {Winter}},\ }\bibfield  {title} {\bibinfo {title} {Quantum state merging and
  negative information},\ }\href@noop {} {\bibfield  {journal} {\bibinfo
  {journal} {Communications in Mathematical Physics}\ }\textbf {\bibinfo
  {volume} {269}},\ \bibinfo {pages} {107} (\bibinfo {year}
  {2007})}\BibitemShut {NoStop}%
\bibitem [{\citenamefont {Arnon-Friedman}\ and\ \citenamefont
  {Bancal}(2019)}]{arnon2019device}%
  \BibitemOpen
  \bibfield  {author} {\bibinfo {author} {\bibfnamefont {R.}~\bibnamefont
  {Arnon-Friedman}}\ and\ \bibinfo {author} {\bibfnamefont {J.-D.}\
  \bibnamefont {Bancal}},\ }\bibfield  {title} {\bibinfo {title}
  {Device-independent certification of one-shot distillable entanglement},\
  }\href@noop {} {\bibfield  {journal} {\bibinfo  {journal} {New Journal of
  Physics}\ }\textbf {\bibinfo {volume} {21}},\ \bibinfo {pages} {033010}
  (\bibinfo {year} {2019})}\BibitemShut {NoStop}%
\bibitem [{\citenamefont {Paw{\l}owski}\ and\ \citenamefont
  {Brunner}(2011)}]{pawlowski2011semi}%
  \BibitemOpen
  \bibfield  {author} {\bibinfo {author} {\bibfnamefont {M.}~\bibnamefont
  {Paw{\l}owski}}\ and\ \bibinfo {author} {\bibfnamefont {N.}~\bibnamefont
  {Brunner}},\ }\bibfield  {title} {\bibinfo {title} {Semi-device-independent
  security of one-way quantum key distribution},\ }\href@noop {} {\bibfield
  {journal} {\bibinfo  {journal} {Physical Review A}\ }\textbf {\bibinfo
  {volume} {84}},\ \bibinfo {pages} {010302} (\bibinfo {year}
  {2011})}\BibitemShut {NoStop}%
\bibitem [{\citenamefont {Pivoluska}\ \emph {et~al.}(2021)\citenamefont
  {Pivoluska}, \citenamefont {Plesch}, \citenamefont {Farkas}, \citenamefont
  {Ru{\v{z}}i{\v{c}}kov{\'a}}, \citenamefont {Flegel}, \citenamefont
  {Valencia}, \citenamefont {McCutcheon}, \citenamefont {Malik},\ and\
  \citenamefont {Aguilar}}]{pivoluska2021semi}%
  \BibitemOpen
  \bibfield  {author} {\bibinfo {author} {\bibfnamefont {M.}~\bibnamefont
  {Pivoluska}}, \bibinfo {author} {\bibfnamefont {M.}~\bibnamefont {Plesch}},
  \bibinfo {author} {\bibfnamefont {M.}~\bibnamefont {Farkas}}, \bibinfo
  {author} {\bibfnamefont {N.}~\bibnamefont {Ru{\v{z}}i{\v{c}}kov{\'a}}},
  \bibinfo {author} {\bibfnamefont {C.}~\bibnamefont {Flegel}}, \bibinfo
  {author} {\bibfnamefont {N.~H.}\ \bibnamefont {Valencia}}, \bibinfo {author}
  {\bibfnamefont {W.}~\bibnamefont {McCutcheon}}, \bibinfo {author}
  {\bibfnamefont {M.}~\bibnamefont {Malik}},\ and\ \bibinfo {author}
  {\bibfnamefont {E.~A.}\ \bibnamefont {Aguilar}},\ }\bibfield  {title}
  {\bibinfo {title} {Semi-device-independent random number generation with
  flexible assumptions},\ }\href@noop {} {\bibfield  {journal} {\bibinfo
  {journal} {npj Quantum Information}\ }\textbf {\bibinfo {volume} {7}},\
  \bibinfo {pages} {50} (\bibinfo {year} {2021})}\BibitemShut {NoStop}%
\bibitem [{\citenamefont {Dorrendorf}\ \emph {et~al.}(2009)\citenamefont
  {Dorrendorf}, \citenamefont {Gutterman},\ and\ \citenamefont
  {Pinkas}}]{dorrendorf2009cryptanalysis}%
  \BibitemOpen
  \bibfield  {author} {\bibinfo {author} {\bibfnamefont {L.}~\bibnamefont
  {Dorrendorf}}, \bibinfo {author} {\bibfnamefont {Z.}~\bibnamefont
  {Gutterman}},\ and\ \bibinfo {author} {\bibfnamefont {B.}~\bibnamefont
  {Pinkas}},\ }\bibfield  {title} {\bibinfo {title} {Cryptanalysis of the
  random number generator of the windows operating system},\ }\href@noop {}
  {\bibfield  {journal} {\bibinfo  {journal} {ACM Transactions on Information
  and System Security (TISSEC)}\ }\textbf {\bibinfo {volume} {13}},\ \bibinfo
  {pages} {1} (\bibinfo {year} {2009})}\BibitemShut {NoStop}%
\bibitem [{\citenamefont {Brunner}\ \emph {et~al.}(2013)\citenamefont
  {Brunner}, \citenamefont {Navascu{\'e}s},\ and\ \citenamefont
  {V{\'e}rtesi}}]{brunner2013dimension}%
  \BibitemOpen
  \bibfield  {author} {\bibinfo {author} {\bibfnamefont {N.}~\bibnamefont
  {Brunner}}, \bibinfo {author} {\bibfnamefont {M.}~\bibnamefont
  {Navascu{\'e}s}},\ and\ \bibinfo {author} {\bibfnamefont {T.}~\bibnamefont
  {V{\'e}rtesi}},\ }\bibfield  {title} {\bibinfo {title} {Dimension witnesses
  and quantum state discrimination},\ }\href@noop {} {\bibfield  {journal}
  {\bibinfo  {journal} {Physical Review Letters}\ }\textbf {\bibinfo {volume}
  {110}},\ \bibinfo {pages} {150501} (\bibinfo {year} {2013})}\BibitemShut
  {NoStop}%
\bibitem [{\citenamefont {Van~Himbeeck}\ \emph {et~al.}(2017)\citenamefont
  {Van~Himbeeck}, \citenamefont {Woodhead}, \citenamefont {Cerf}, \citenamefont
  {Garc{\'\i}a-Patr{\'o}n},\ and\ \citenamefont {Pironio}}]{van2017semi}%
  \BibitemOpen
  \bibfield  {author} {\bibinfo {author} {\bibfnamefont {T.}~\bibnamefont
  {Van~Himbeeck}}, \bibinfo {author} {\bibfnamefont {E.}~\bibnamefont
  {Woodhead}}, \bibinfo {author} {\bibfnamefont {N.~J.}\ \bibnamefont {Cerf}},
  \bibinfo {author} {\bibfnamefont {R.}~\bibnamefont
  {Garc{\'\i}a-Patr{\'o}n}},\ and\ \bibinfo {author} {\bibfnamefont
  {S.}~\bibnamefont {Pironio}},\ }\bibfield  {title} {\bibinfo {title}
  {Semi-device-independent framework based on natural physical assumptions},\
  }\href@noop {} {\bibfield  {journal} {\bibinfo  {journal} {Quantum}\ }\textbf
  {\bibinfo {volume} {1}},\ \bibinfo {pages} {33} (\bibinfo {year}
  {2017})}\BibitemShut {NoStop}%
\bibitem [{\citenamefont
  {Chatzidimitriou-Dreismann}(2020)}]{chatzidimitriou2020experimental}%
  \BibitemOpen
  \bibfield  {author} {\bibinfo {author} {\bibfnamefont {C.~A.}\ \bibnamefont
  {Chatzidimitriou-Dreismann}},\ }\bibfield  {title} {\bibinfo {title}
  {Experimental implications of negative quantum conditional
  entropy—$\text{{H}}_2$ mobility in nanoporous materials},\ }\href@noop {}
  {\bibfield  {journal} {\bibinfo  {journal} {Applied Sciences}\ }\textbf
  {\bibinfo {volume} {10}},\ \bibinfo {pages} {8266} (\bibinfo {year}
  {2020})}\BibitemShut {NoStop}%
\bibitem [{\citenamefont {Sheshadri}\ and\ \citenamefont
  {Chainani}(2021)}]{sheshadri2021unconventional}%
  \BibitemOpen
  \bibfield  {author} {\bibinfo {author} {\bibfnamefont {K.}~\bibnamefont
  {Sheshadri}}\ and\ \bibinfo {author} {\bibfnamefont {A.}~\bibnamefont
  {Chainani}},\ }\bibfield  {title} {\bibinfo {title} {Unconventional
  superfluidity in a model of {F}ermi-{B}ose mixtures},\ }\href@noop {}
  {\bibfield  {journal} {\bibinfo  {journal} {arXiv preprint arXiv:2101.08513}\
  } (\bibinfo {year} {2021})}\BibitemShut {NoStop}%
\bibitem [{\citenamefont {Vandenberghe}\ and\ \citenamefont
  {Boyd}(1996)}]{sdp}%
  \BibitemOpen
  \bibfield  {author} {\bibinfo {author} {\bibfnamefont {L.}~\bibnamefont
  {Vandenberghe}}\ and\ \bibinfo {author} {\bibfnamefont {S.}~\bibnamefont
  {Boyd}},\ }\bibfield  {title} {\bibinfo {title} {Semidefinite programming},\
  }\href@noop {} {\bibfield  {journal} {\bibinfo  {journal} {SIAM review}\
  }\textbf {\bibinfo {volume} {38}},\ \bibinfo {pages} {49} (\bibinfo {year}
  {1996})}\BibitemShut {NoStop}%
\bibitem [{\citenamefont {Skrzypczyk}\ and\ \citenamefont
  {Cavalcanti}(2023)}]{Skrzypczyk2023}%
  \BibitemOpen
  \bibfield  {author} {\bibinfo {author} {\bibfnamefont {P.}~\bibnamefont
  {Skrzypczyk}}\ and\ \bibinfo {author} {\bibfnamefont {D.}~\bibnamefont
  {Cavalcanti}},\ }\href {https://doi.org/10.1088/978-0-7503-3343-6} {\emph
  {\bibinfo {title} {{Semidefinite Programming in Quantum Information
  Science}}}},\ 2053-2563\ (\bibinfo  {publisher} {IOP Publishing},\ \bibinfo
  {year} {2023})\BibitemShut {NoStop}%
\bibitem [{\citenamefont {Mironowicz}(2024)}]{mironowicz2024semi}%
  \BibitemOpen
  \bibfield  {author} {\bibinfo {author} {\bibfnamefont {P.}~\bibnamefont
  {Mironowicz}},\ }\bibfield  {title} {\bibinfo {title} {Semi-definite
  programming and quantum information},\ }\href@noop {} {\bibfield  {journal}
  {\bibinfo  {journal} {Journal of Physics A: Mathematical and Theoretical}\
  }\textbf {\bibinfo {volume} {57}},\ \bibinfo {pages} {163002} (\bibinfo
  {year} {2024})}\BibitemShut {NoStop}%
\bibitem [{\citenamefont {Tavakoli}\ \emph {et~al.}(2023)\citenamefont
  {Tavakoli}, \citenamefont {Pozas-Kerstjens}, \citenamefont {Brown},\ and\
  \citenamefont {Ara{\'u}jo}}]{tavakoli2023semidefinite}%
  \BibitemOpen
  \bibfield  {author} {\bibinfo {author} {\bibfnamefont {A.}~\bibnamefont
  {Tavakoli}}, \bibinfo {author} {\bibfnamefont {A.}~\bibnamefont
  {Pozas-Kerstjens}}, \bibinfo {author} {\bibfnamefont {P.}~\bibnamefont
  {Brown}},\ and\ \bibinfo {author} {\bibfnamefont {M.}~\bibnamefont
  {Ara{\'u}jo}},\ }\bibfield  {title} {\bibinfo {title} {Semidefinite
  programming relaxations for quantum correlations},\ }\href@noop {} {\bibfield
   {journal} {\bibinfo  {journal} {arXiv:2307.02551}\ } (\bibinfo {year}
  {2023})}\BibitemShut {NoStop}%
\bibitem [{\citenamefont {P{\'a}l}\ and\ \citenamefont
  {V{\'e}rtesi}(2010)}]{pal2010maximal}%
  \BibitemOpen
  \bibfield  {author} {\bibinfo {author} {\bibfnamefont {K.~F.}\ \bibnamefont
  {P{\'a}l}}\ and\ \bibinfo {author} {\bibfnamefont {T.}~\bibnamefont
  {V{\'e}rtesi}},\ }\bibfield  {title} {\bibinfo {title} {Maximal violation of
  a bipartite three-setting, two-outcome {B}ell inequality using
  infinite-dimensional quantum systems},\ }\href@noop {} {\bibfield  {journal}
  {\bibinfo  {journal} {Physical Review A}\ }\textbf {\bibinfo {volume} {82}},\
  \bibinfo {pages} {022116} (\bibinfo {year} {2010})}\BibitemShut {NoStop}%
\bibitem [{\citenamefont {Fawzi}\ and\ \citenamefont
  {Fawzi}(2018)}]{fawzi2018efficient}%
  \BibitemOpen
  \bibfield  {author} {\bibinfo {author} {\bibfnamefont {H.}~\bibnamefont
  {Fawzi}}\ and\ \bibinfo {author} {\bibfnamefont {O.}~\bibnamefont {Fawzi}},\
  }\bibfield  {title} {\bibinfo {title} {Efficient optimization of the quantum
  relative entropy},\ }\href@noop {} {\bibfield  {journal} {\bibinfo  {journal}
  {Journal of Physics A: Mathematical and Theoretical}\ }\textbf {\bibinfo
  {volume} {51}},\ \bibinfo {pages} {154003} (\bibinfo {year}
  {2018})}\BibitemShut {NoStop}%
\bibitem [{\citenamefont {Fawzi}\ \emph {et~al.}(2019)\citenamefont {Fawzi},
  \citenamefont {Saunderson},\ and\ \citenamefont {Parrilo}}]{logapx}%
  \BibitemOpen
  \bibfield  {author} {\bibinfo {author} {\bibfnamefont {H.}~\bibnamefont
  {Fawzi}}, \bibinfo {author} {\bibfnamefont {J.}~\bibnamefont {Saunderson}},\
  and\ \bibinfo {author} {\bibfnamefont {P.~A.}\ \bibnamefont {Parrilo}},\
  }\bibfield  {title} {\bibinfo {title} {Semidefinite approximations of the
  matrix logarithm},\ }\href@noop {} {\bibfield  {journal} {\bibinfo  {journal}
  {Foundations of Computational Mathematics}\ }\textbf {\bibinfo {volume}
  {19}},\ \bibinfo {pages} {259} (\bibinfo {year} {2019})}\BibitemShut
  {NoStop}%
\bibitem [{\citenamefont {Patro}\ \emph {et~al.}(2017)\citenamefont {Patro},
  \citenamefont {Chakrabarty},\ and\ \citenamefont {Ganguly}}]{patro2017non}%
  \BibitemOpen
  \bibfield  {author} {\bibinfo {author} {\bibfnamefont {S.}~\bibnamefont
  {Patro}}, \bibinfo {author} {\bibfnamefont {I.}~\bibnamefont {Chakrabarty}},\
  and\ \bibinfo {author} {\bibfnamefont {N.}~\bibnamefont {Ganguly}},\
  }\bibfield  {title} {\bibinfo {title} {Non-negativity of conditional von
  {N}eumann entropy and global unitary operations},\ }\href@noop {} {\bibfield
  {journal} {\bibinfo  {journal} {Physical Review A}\ }\textbf {\bibinfo
  {volume} {96}},\ \bibinfo {pages} {062102} (\bibinfo {year}
  {2017})}\BibitemShut {NoStop}%
\bibitem [{\citenamefont {Luc}(2020)}]{luc2020quantumness}%
  \BibitemOpen
  \bibfield  {author} {\bibinfo {author} {\bibfnamefont {J.}~\bibnamefont
  {Luc}},\ }\bibfield  {title} {\bibinfo {title} {Quantumness of states and
  unitary operations},\ }\href@noop {} {\bibfield  {journal} {\bibinfo
  {journal} {Foundations of Physics}\ }\textbf {\bibinfo {volume} {50}},\
  \bibinfo {pages} {1645} (\bibinfo {year} {2020})}\BibitemShut {NoStop}%
\bibitem [{\citenamefont {Kumar}\ and\ \citenamefont
  {Ganguly}(2023)}]{kumar2023quantum}%
  \BibitemOpen
  \bibfield  {author} {\bibinfo {author} {\bibfnamefont {K.}~\bibnamefont
  {Kumar}}\ and\ \bibinfo {author} {\bibfnamefont {N.}~\bibnamefont
  {Ganguly}},\ }\bibfield  {title} {\bibinfo {title} {Quantum conditional
  entropies and steerability of states with maximally mixed marginals},\
  }\href@noop {} {\bibfield  {journal} {\bibinfo  {journal} {Physical Review
  A}\ }\textbf {\bibinfo {volume} {107}},\ \bibinfo {pages} {032206} (\bibinfo
  {year} {2023})}\BibitemShut {NoStop}%
\bibitem [{\citenamefont {Friis}\ \emph {et~al.}(2017)\citenamefont {Friis},
  \citenamefont {Bulusu},\ and\ \citenamefont {Bertlmann}}]{friis2017geometry}%
  \BibitemOpen
  \bibfield  {author} {\bibinfo {author} {\bibfnamefont {N.}~\bibnamefont
  {Friis}}, \bibinfo {author} {\bibfnamefont {S.}~\bibnamefont {Bulusu}},\ and\
  \bibinfo {author} {\bibfnamefont {R.~A.}\ \bibnamefont {Bertlmann}},\
  }\bibfield  {title} {\bibinfo {title} {Geometry of two-qubit states with
  negative conditional entropy},\ }\href@noop {} {\bibfield  {journal}
  {\bibinfo  {journal} {Journal of Physics A: Mathematical and Theoretical}\
  }\textbf {\bibinfo {volume} {50}},\ \bibinfo {pages} {125301} (\bibinfo
  {year} {2017})}\BibitemShut {NoStop}%
\bibitem [{\citenamefont {Vempati}\ \emph {et~al.}(2021)\citenamefont
  {Vempati}, \citenamefont {Ganguly}, \citenamefont {Chakrabarty},\ and\
  \citenamefont {Pati}}]{witnessing}%
  \BibitemOpen
  \bibfield  {author} {\bibinfo {author} {\bibfnamefont {M.}~\bibnamefont
  {Vempati}}, \bibinfo {author} {\bibfnamefont {N.}~\bibnamefont {Ganguly}},
  \bibinfo {author} {\bibfnamefont {I.}~\bibnamefont {Chakrabarty}},\ and\
  \bibinfo {author} {\bibfnamefont {A.~K.}\ \bibnamefont {Pati}},\ }\bibfield
  {title} {\bibinfo {title} {Witnessing negative conditional entropy},\
  }\href@noop {} {\bibfield  {journal} {\bibinfo  {journal} {Physical Review
  A}\ }\textbf {\bibinfo {volume} {104}},\ \bibinfo {pages} {012417} (\bibinfo
  {year} {2021})}\BibitemShut {NoStop}%
\bibitem [{\citenamefont {Vempati}\ \emph {et~al.}(2022)\citenamefont
  {Vempati}, \citenamefont {Shah}, \citenamefont {Ganguly},\ and\ \citenamefont
  {Chakrabarty}}]{vempati2022unital}%
  \BibitemOpen
  \bibfield  {author} {\bibinfo {author} {\bibfnamefont {M.}~\bibnamefont
  {Vempati}}, \bibinfo {author} {\bibfnamefont {S.}~\bibnamefont {Shah}},
  \bibinfo {author} {\bibfnamefont {N.}~\bibnamefont {Ganguly}},\ and\ \bibinfo
  {author} {\bibfnamefont {I.}~\bibnamefont {Chakrabarty}},\ }\bibfield
  {title} {\bibinfo {title} {A-unital operations and quantum conditional
  entropy},\ }\href@noop {} {\bibfield  {journal} {\bibinfo  {journal}
  {Quantum}\ }\textbf {\bibinfo {volume} {6}},\ \bibinfo {pages} {641}
  (\bibinfo {year} {2022})}\BibitemShut {NoStop}%
\bibitem [{\citenamefont {Clauser}\ \emph {et~al.}(1969)\citenamefont
  {Clauser}, \citenamefont {Horne}, \citenamefont {Shimony},\ and\
  \citenamefont {Holt}}]{clauser1969proposed}%
  \BibitemOpen
  \bibfield  {author} {\bibinfo {author} {\bibfnamefont {J.~F.}\ \bibnamefont
  {Clauser}}, \bibinfo {author} {\bibfnamefont {M.~A.}\ \bibnamefont {Horne}},
  \bibinfo {author} {\bibfnamefont {A.}~\bibnamefont {Shimony}},\ and\ \bibinfo
  {author} {\bibfnamefont {R.~A.}\ \bibnamefont {Holt}},\ }\bibfield  {title}
  {\bibinfo {title} {Proposed experiment to test local hidden-variable
  theories},\ }\href@noop {} {\bibfield  {journal} {\bibinfo  {journal}
  {Physical Review Letters}\ }\textbf {\bibinfo {volume} {23}},\ \bibinfo
  {pages} {880} (\bibinfo {year} {1969})}\BibitemShut {NoStop}%
\bibitem [{\citenamefont {Masanes}\ \emph {et~al.}(2011)\citenamefont
  {Masanes}, \citenamefont {Pironio},\ and\ \citenamefont
  {Ac{\'\i}n}}]{masanes2011secure}%
  \BibitemOpen
  \bibfield  {author} {\bibinfo {author} {\bibfnamefont {L.}~\bibnamefont
  {Masanes}}, \bibinfo {author} {\bibfnamefont {S.}~\bibnamefont {Pironio}},\
  and\ \bibinfo {author} {\bibfnamefont {A.}~\bibnamefont {Ac{\'\i}n}},\
  }\bibfield  {title} {\bibinfo {title} {Secure device-independent quantum key
  distribution with causally independent measurement devices},\ }\href@noop {}
  {\bibfield  {journal} {\bibinfo  {journal} {Nature communications}\ }\textbf
  {\bibinfo {volume} {2}},\ \bibinfo {pages} {238} (\bibinfo {year}
  {2011})}\BibitemShut {NoStop}%
\bibitem [{\citenamefont {Karczewski}\ \emph {et~al.}(2022)\citenamefont
  {Karczewski}, \citenamefont {Scala}, \citenamefont {Mandarino}, \citenamefont
  {Sainz},\ and\ \citenamefont {{\.Z}ukowski}}]{karczewski2022avenues}%
  \BibitemOpen
  \bibfield  {author} {\bibinfo {author} {\bibfnamefont {M.}~\bibnamefont
  {Karczewski}}, \bibinfo {author} {\bibfnamefont {G.}~\bibnamefont {Scala}},
  \bibinfo {author} {\bibfnamefont {A.}~\bibnamefont {Mandarino}}, \bibinfo
  {author} {\bibfnamefont {A.~B.}\ \bibnamefont {Sainz}},\ and\ \bibinfo
  {author} {\bibfnamefont {M.}~\bibnamefont {{\.Z}ukowski}},\ }\bibfield
  {title} {\bibinfo {title} {Avenues to generalising {B}ell inequalities},\
  }\href@noop {} {\bibfield  {journal} {\bibinfo  {journal} {Journal of Physics
  A: Mathematical and Theoretical}\ }\textbf {\bibinfo {volume} {55}},\
  \bibinfo {pages} {384011} (\bibinfo {year} {2022})}\BibitemShut {NoStop}%
\bibitem [{\citenamefont {Barrett}\ \emph {et~al.}(2005)\citenamefont
  {Barrett}, \citenamefont {Linden}, \citenamefont {Massar}, \citenamefont
  {Pironio}, \citenamefont {Popescu},\ and\ \citenamefont
  {Roberts}}]{barrett2005nonlocal}%
  \BibitemOpen
  \bibfield  {author} {\bibinfo {author} {\bibfnamefont {J.}~\bibnamefont
  {Barrett}}, \bibinfo {author} {\bibfnamefont {N.}~\bibnamefont {Linden}},
  \bibinfo {author} {\bibfnamefont {S.}~\bibnamefont {Massar}}, \bibinfo
  {author} {\bibfnamefont {S.}~\bibnamefont {Pironio}}, \bibinfo {author}
  {\bibfnamefont {S.}~\bibnamefont {Popescu}},\ and\ \bibinfo {author}
  {\bibfnamefont {D.}~\bibnamefont {Roberts}},\ }\bibfield  {title} {\bibinfo
  {title} {Nonlocal correlations as an information-theoretic resource},\
  }\href@noop {} {\bibfield  {journal} {\bibinfo  {journal} {Physical Review
  A}\ }\textbf {\bibinfo {volume} {71}},\ \bibinfo {pages} {022101} (\bibinfo
  {year} {2005})}\BibitemShut {NoStop}%
\bibitem [{\citenamefont {Horodecki}\ and\ \citenamefont
  {Oppenheim}(2013)}]{horodecki2013quantumness}%
  \BibitemOpen
  \bibfield  {author} {\bibinfo {author} {\bibfnamefont {M.}~\bibnamefont
  {Horodecki}}\ and\ \bibinfo {author} {\bibfnamefont {J.}~\bibnamefont
  {Oppenheim}},\ }\bibfield  {title} {\bibinfo {title} {{(Quantumness in the
  context of) resource theories}},\ }\href@noop {} {\bibfield  {journal}
  {\bibinfo  {journal} {International Journal of Modern Physics B}\ }\textbf
  {\bibinfo {volume} {27}},\ \bibinfo {pages} {1345019} (\bibinfo {year}
  {2013})}\BibitemShut {NoStop}%
\bibitem [{\citenamefont {Werner}(1989)}]{werner1989quantum}%
  \BibitemOpen
  \bibfield  {author} {\bibinfo {author} {\bibfnamefont {R.~F.}\ \bibnamefont
  {Werner}},\ }\bibfield  {title} {\bibinfo {title} {Quantum states with
  {Einstein-Podolsky-Rosen} correlations admitting a hidden-variable model},\
  }\href@noop {} {\bibfield  {journal} {\bibinfo  {journal} {Physical Review
  A}\ }\textbf {\bibinfo {volume} {40}},\ \bibinfo {pages} {4277} (\bibinfo
  {year} {1989})}\BibitemShut {NoStop}%
\bibitem [{\citenamefont {Wilms}\ \emph {et~al.}(2008)\citenamefont {Wilms},
  \citenamefont {Disser}, \citenamefont {Alber},\ and\ \citenamefont
  {Percival}}]{wilms2008local}%
  \BibitemOpen
  \bibfield  {author} {\bibinfo {author} {\bibfnamefont {J.}~\bibnamefont
  {Wilms}}, \bibinfo {author} {\bibfnamefont {Y.}~\bibnamefont {Disser}},
  \bibinfo {author} {\bibfnamefont {G.}~\bibnamefont {Alber}},\ and\ \bibinfo
  {author} {\bibfnamefont {I.~C.}\ \bibnamefont {Percival}},\ }\bibfield
  {title} {\bibinfo {title} {Local realism, detection efficiencies, and
  probability polytopes},\ }\href@noop {} {\bibfield  {journal} {\bibinfo
  {journal} {Physical Review A}\ }\textbf {\bibinfo {volume} {78}},\ \bibinfo
  {pages} {032116} (\bibinfo {year} {2008})}\BibitemShut {NoStop}%
\bibitem [{\citenamefont {Czechlewski}\ and\ \citenamefont
  {Paw{\l}owski}(2018)}]{czechlewski2018influence}%
  \BibitemOpen
  \bibfield  {author} {\bibinfo {author} {\bibfnamefont {M.}~\bibnamefont
  {Czechlewski}}\ and\ \bibinfo {author} {\bibfnamefont {M.}~\bibnamefont
  {Paw{\l}owski}},\ }\bibfield  {title} {\bibinfo {title} {Influence of the
  choice of postprocessing method on {B}ell inequalities},\ }\href@noop {}
  {\bibfield  {journal} {\bibinfo  {journal} {Physical Review A}\ }\textbf
  {\bibinfo {volume} {97}},\ \bibinfo {pages} {062123} (\bibinfo {year}
  {2018})}\BibitemShut {NoStop}%
\bibitem [{\citenamefont {{\v{S}}upi{\'c}}\ and\ \citenamefont
  {Bowles}(2020)}]{vsupic2020self}%
  \BibitemOpen
  \bibfield  {author} {\bibinfo {author} {\bibfnamefont {I.}~\bibnamefont
  {{\v{S}}upi{\'c}}}\ and\ \bibinfo {author} {\bibfnamefont {J.}~\bibnamefont
  {Bowles}},\ }\bibfield  {title} {\bibinfo {title} {Self-testing of quantum
  systems: a review},\ }\href@noop {} {\bibfield  {journal} {\bibinfo
  {journal} {Quantum}\ }\textbf {\bibinfo {volume} {4}},\ \bibinfo {pages}
  {337} (\bibinfo {year} {2020})}\BibitemShut {NoStop}%
\bibitem [{\citenamefont {Summers}\ and\ \citenamefont
  {Werner}(1987)}]{summers1987maximal}%
  \BibitemOpen
  \bibfield  {author} {\bibinfo {author} {\bibfnamefont {S.~J.}\ \bibnamefont
  {Summers}}\ and\ \bibinfo {author} {\bibfnamefont {R.}~\bibnamefont
  {Werner}},\ }\bibfield  {title} {\bibinfo {title} {Maximal violation of
  {B}ell's inequalities is generic in quantum field theory},\ }\href@noop {}
  {\bibfield  {journal} {\bibinfo  {journal} {Communications in Mathematical
  Physics}\ }\textbf {\bibinfo {volume} {110}},\ \bibinfo {pages} {247}
  (\bibinfo {year} {1987})}\BibitemShut {NoStop}%
\bibitem [{\citenamefont {Popescu}\ and\ \citenamefont
  {Rohrlich}(1992)}]{popescu1992states}%
  \BibitemOpen
  \bibfield  {author} {\bibinfo {author} {\bibfnamefont {S.}~\bibnamefont
  {Popescu}}\ and\ \bibinfo {author} {\bibfnamefont {D.}~\bibnamefont
  {Rohrlich}},\ }\bibfield  {title} {\bibinfo {title} {Which states violate
  {B}ell's inequality maximally?},\ }\href@noop {} {\bibfield  {journal}
  {\bibinfo  {journal} {Physics Letters A}\ }\textbf {\bibinfo {volume}
  {169}},\ \bibinfo {pages} {411} (\bibinfo {year} {1992})}\BibitemShut
  {NoStop}%
\bibitem [{\citenamefont {Braunstein}\ \emph {et~al.}(1992)\citenamefont
  {Braunstein}, \citenamefont {Mann},\ and\ \citenamefont
  {Revzen}}]{braunstein1992maximal}%
  \BibitemOpen
  \bibfield  {author} {\bibinfo {author} {\bibfnamefont {S.~L.}\ \bibnamefont
  {Braunstein}}, \bibinfo {author} {\bibfnamefont {A.}~\bibnamefont {Mann}},\
  and\ \bibinfo {author} {\bibfnamefont {M.}~\bibnamefont {Revzen}},\
  }\bibfield  {title} {\bibinfo {title} {Maximal violation of {B}ell
  inequalities for mixed states},\ }\href@noop {} {\bibfield  {journal}
  {\bibinfo  {journal} {Physical Review Letters}\ }\textbf {\bibinfo {volume}
  {68}},\ \bibinfo {pages} {3259} (\bibinfo {year} {1992})}\BibitemShut
  {NoStop}%
\bibitem [{\citenamefont {Tsirelson}(1993)}]{tsirelson1993some}%
  \BibitemOpen
  \bibfield  {author} {\bibinfo {author} {\bibfnamefont {B.~S.}\ \bibnamefont
  {Tsirelson}},\ }\bibfield  {title} {\bibinfo {title} {Some results and
  problems on quantum {B}ell-type inequalities},\ }\href@noop {} {\bibfield
  {journal} {\bibinfo  {journal} {Hadronic Journal Supplement}\ }\textbf
  {\bibinfo {volume} {8}},\ \bibinfo {pages} {329} (\bibinfo {year}
  {1993})}\BibitemShut {NoStop}%
\bibitem [{\citenamefont {Slofstra}(2011)}]{slofstra2011lower}%
  \BibitemOpen
  \bibfield  {author} {\bibinfo {author} {\bibfnamefont {W.}~\bibnamefont
  {Slofstra}},\ }\bibfield  {title} {\bibinfo {title} {Lower bounds on the
  entanglement needed to play {XOR} non-local games},\ }\href@noop {}
  {\bibfield  {journal} {\bibinfo  {journal} {Journal of Mathematical Physics}\
  }\textbf {\bibinfo {volume} {52}} (\bibinfo {year} {2011})}\BibitemShut
  {NoStop}%
\bibitem [{\citenamefont {Miller}\ and\ \citenamefont
  {Shi}(2013)}]{miller2013optimal}%
  \BibitemOpen
  \bibfield  {author} {\bibinfo {author} {\bibfnamefont {C.~A.}\ \bibnamefont
  {Miller}}\ and\ \bibinfo {author} {\bibfnamefont {Y.}~\bibnamefont {Shi}},\
  }\bibfield  {title} {\bibinfo {title} {Optimal robust self-testing by binary
  nonlocal {XOR} games},\ }in\ \href@noop {} {\emph {\bibinfo {booktitle} {8th
  Conference on the Theory of Quantum Computation, Communication and
  Cryptography (TQC 2013)}}}\ (\bibinfo {organization} {Schloss
  Dagstuhl-Leibniz-Zentrum fuer Informatik},\ \bibinfo {year}
  {2013})\BibitemShut {NoStop}%
\bibitem [{\citenamefont {Ramanathan}\ \emph {et~al.}(2018)\citenamefont
  {Ramanathan}, \citenamefont {Goyeneche}, \citenamefont {Muhammad},
  \citenamefont {Mironowicz}, \citenamefont {Gr{\"u}nfeld}, \citenamefont
  {Bourennane},\ and\ \citenamefont {Horodecki}}]{ramanathan2018steering}%
  \BibitemOpen
  \bibfield  {author} {\bibinfo {author} {\bibfnamefont {R.}~\bibnamefont
  {Ramanathan}}, \bibinfo {author} {\bibfnamefont {D.}~\bibnamefont
  {Goyeneche}}, \bibinfo {author} {\bibfnamefont {S.}~\bibnamefont {Muhammad}},
  \bibinfo {author} {\bibfnamefont {P.}~\bibnamefont {Mironowicz}}, \bibinfo
  {author} {\bibfnamefont {M.}~\bibnamefont {Gr{\"u}nfeld}}, \bibinfo {author}
  {\bibfnamefont {M.}~\bibnamefont {Bourennane}},\ and\ \bibinfo {author}
  {\bibfnamefont {P.}~\bibnamefont {Horodecki}},\ }\bibfield  {title} {\bibinfo
  {title} {Steering is an essential feature of non-locality in quantum
  theory},\ }\href@noop {} {\bibfield  {journal} {\bibinfo  {journal} {Nature
  communications}\ }\textbf {\bibinfo {volume} {9}},\ \bibinfo {pages} {4244}
  (\bibinfo {year} {2018})}\BibitemShut {NoStop}%
\bibitem [{\citenamefont {Mironowicz}\ and\ \citenamefont
  {Paw{\l}owski}(2019)}]{mironowicz2019experimentally}%
  \BibitemOpen
  \bibfield  {author} {\bibinfo {author} {\bibfnamefont {P.}~\bibnamefont
  {Mironowicz}}\ and\ \bibinfo {author} {\bibfnamefont {M.}~\bibnamefont
  {Paw{\l}owski}},\ }\bibfield  {title} {\bibinfo {title} {Experimentally
  feasible semi-device-independent certification of four-outcome
  positive-operator-valued measurements},\ }\href@noop {} {\bibfield  {journal}
  {\bibinfo  {journal} {Physical Review A}\ }\textbf {\bibinfo {volume}
  {100}},\ \bibinfo {pages} {030301} (\bibinfo {year} {2019})}\BibitemShut
  {NoStop}%
\bibitem [{\citenamefont {Valcarce}\ \emph {et~al.}(2020)\citenamefont
  {Valcarce}, \citenamefont {Sekatski}, \citenamefont {Orsucci}, \citenamefont
  {Oudot}, \citenamefont {Bancal},\ and\ \citenamefont
  {Sangouard}}]{valcarce2020minimum}%
  \BibitemOpen
  \bibfield  {author} {\bibinfo {author} {\bibfnamefont {X.}~\bibnamefont
  {Valcarce}}, \bibinfo {author} {\bibfnamefont {P.}~\bibnamefont {Sekatski}},
  \bibinfo {author} {\bibfnamefont {D.}~\bibnamefont {Orsucci}}, \bibinfo
  {author} {\bibfnamefont {E.}~\bibnamefont {Oudot}}, \bibinfo {author}
  {\bibfnamefont {J.-D.}\ \bibnamefont {Bancal}},\ and\ \bibinfo {author}
  {\bibfnamefont {N.}~\bibnamefont {Sangouard}},\ }\bibfield  {title} {\bibinfo
  {title} {What is the minimum {CHSH} score certifying that a state resembles
  the singlet?},\ }\href@noop {} {\bibfield  {journal} {\bibinfo  {journal}
  {Quantum}\ }\textbf {\bibinfo {volume} {4}},\ \bibinfo {pages} {246}
  (\bibinfo {year} {2020})}\BibitemShut {NoStop}%
\bibitem [{\citenamefont {Smania}\ \emph {et~al.}(2020)\citenamefont {Smania},
  \citenamefont {Mironowicz}, \citenamefont {Nawareg}, \citenamefont
  {Paw{\l}owski}, \citenamefont {Cabello},\ and\ \citenamefont
  {Bourennane}}]{smania2020experimental}%
  \BibitemOpen
  \bibfield  {author} {\bibinfo {author} {\bibfnamefont {M.}~\bibnamefont
  {Smania}}, \bibinfo {author} {\bibfnamefont {P.}~\bibnamefont {Mironowicz}},
  \bibinfo {author} {\bibfnamefont {M.}~\bibnamefont {Nawareg}}, \bibinfo
  {author} {\bibfnamefont {M.}~\bibnamefont {Paw{\l}owski}}, \bibinfo {author}
  {\bibfnamefont {A.}~\bibnamefont {Cabello}},\ and\ \bibinfo {author}
  {\bibfnamefont {M.}~\bibnamefont {Bourennane}},\ }\bibfield  {title}
  {\bibinfo {title} {Experimental certification of an informationally complete
  quantum measurement in a device-independent protocol},\ }\href@noop {}
  {\bibfield  {journal} {\bibinfo  {journal} {Optica}\ }\textbf {\bibinfo
  {volume} {7}},\ \bibinfo {pages} {123} (\bibinfo {year} {2020})}\BibitemShut
  {NoStop}%
\bibitem [{\citenamefont {Valcarce}\ \emph {et~al.}(2022)\citenamefont
  {Valcarce}, \citenamefont {Zivy}, \citenamefont {Sangouard},\ and\
  \citenamefont {Sekatski}}]{valcarce2022self}%
  \BibitemOpen
  \bibfield  {author} {\bibinfo {author} {\bibfnamefont {X.}~\bibnamefont
  {Valcarce}}, \bibinfo {author} {\bibfnamefont {J.}~\bibnamefont {Zivy}},
  \bibinfo {author} {\bibfnamefont {N.}~\bibnamefont {Sangouard}},\ and\
  \bibinfo {author} {\bibfnamefont {P.}~\bibnamefont {Sekatski}},\ }\bibfield
  {title} {\bibinfo {title} {Self-testing two-qubit maximally entangled states
  from generalized {C}lauser-{H}orne-{S}himony-{H}olt tests},\ }\href@noop {}
  {\bibfield  {journal} {\bibinfo  {journal} {Physical Review Research}\
  }\textbf {\bibinfo {volume} {4}},\ \bibinfo {pages} {013049} (\bibinfo {year}
  {2022})}\BibitemShut {NoStop}%
\bibitem [{\citenamefont {Braunstein}\ and\ \citenamefont
  {Caves}(1988)}]{braunstein1988information}%
  \BibitemOpen
  \bibfield  {author} {\bibinfo {author} {\bibfnamefont {S.~L.}\ \bibnamefont
  {Braunstein}}\ and\ \bibinfo {author} {\bibfnamefont {C.~M.}\ \bibnamefont
  {Caves}},\ }\bibfield  {title} {\bibinfo {title} {Information-theoretic
  {B}ell inequalities},\ }\href@noop {} {\bibfield  {journal} {\bibinfo
  {journal} {Physical Review Letters}\ }\textbf {\bibinfo {volume} {61}},\
  \bibinfo {pages} {662} (\bibinfo {year} {1988})}\BibitemShut {NoStop}%
\bibitem [{\citenamefont {{\v{S}}upi{\'c}}\ \emph {et~al.}(2016)\citenamefont
  {{\v{S}}upi{\'c}}, \citenamefont {Augusiak}, \citenamefont {Salavrakos},\
  and\ \citenamefont {Ac{\'\i}n}}]{vsupic2016self}%
  \BibitemOpen
  \bibfield  {author} {\bibinfo {author} {\bibfnamefont {I.}~\bibnamefont
  {{\v{S}}upi{\'c}}}, \bibinfo {author} {\bibfnamefont {R.}~\bibnamefont
  {Augusiak}}, \bibinfo {author} {\bibfnamefont {A.}~\bibnamefont
  {Salavrakos}},\ and\ \bibinfo {author} {\bibfnamefont {A.}~\bibnamefont
  {Ac{\'\i}n}},\ }\bibfield  {title} {\bibinfo {title} {Self-testing protocols
  based on the chained {B}ell inequalities},\ }\href@noop {} {\bibfield
  {journal} {\bibinfo  {journal} {New Journal of Physics}\ }\textbf {\bibinfo
  {volume} {18}},\ \bibinfo {pages} {035013} (\bibinfo {year}
  {2016})}\BibitemShut {NoStop}%
\bibitem [{\citenamefont {Mironowicz}\ and\ \citenamefont
  {Paw{\l}owski}(2013)}]{PRA.88.032319}%
  \BibitemOpen
  \bibfield  {author} {\bibinfo {author} {\bibfnamefont {P.}~\bibnamefont
  {Mironowicz}}\ and\ \bibinfo {author} {\bibfnamefont {M.}~\bibnamefont
  {Paw{\l}owski}},\ }\bibfield  {title} {\bibinfo {title} {Robustness of
  quantum-randomness expansion protocols in the presence of noise},\
  }\href@noop {} {\bibfield  {journal} {\bibinfo  {journal} {Physical Review
  A}\ }\textbf {\bibinfo {volume} {88}},\ \bibinfo {pages} {032319} (\bibinfo
  {year} {2013})}\BibitemShut {NoStop}%
\bibitem [{\citenamefont {Brown}\ \emph {et~al.}(2019)\citenamefont {Brown},
  \citenamefont {Ragy},\ and\ \citenamefont {Colbeck}}]{brown2019framework}%
  \BibitemOpen
  \bibfield  {author} {\bibinfo {author} {\bibfnamefont {P.~J.}\ \bibnamefont
  {Brown}}, \bibinfo {author} {\bibfnamefont {S.}~\bibnamefont {Ragy}},\ and\
  \bibinfo {author} {\bibfnamefont {R.}~\bibnamefont {Colbeck}},\ }\bibfield
  {title} {\bibinfo {title} {A framework for quantum-secure device-independent
  randomness expansion},\ }\href@noop {} {\bibfield  {journal} {\bibinfo
  {journal} {IEEE Transactions on Information Theory}\ }\textbf {\bibinfo
  {volume} {66}},\ \bibinfo {pages} {2964} (\bibinfo {year}
  {2019})}\BibitemShut {NoStop}%
\bibitem [{\citenamefont {Xiao}\ \emph {et~al.}(2023)\citenamefont {Xiao},
  \citenamefont {Li}, \citenamefont {Wang}, \citenamefont {Li},\ and\
  \citenamefont {Fei}}]{xiao2023device}%
  \BibitemOpen
  \bibfield  {author} {\bibinfo {author} {\bibfnamefont {Y.}~\bibnamefont
  {Xiao}}, \bibinfo {author} {\bibfnamefont {X.}~\bibnamefont {Li}}, \bibinfo
  {author} {\bibfnamefont {J.}~\bibnamefont {Wang}}, \bibinfo {author}
  {\bibfnamefont {M.}~\bibnamefont {Li}},\ and\ \bibinfo {author}
  {\bibfnamefont {S.-M.}\ \bibnamefont {Fei}},\ }\bibfield  {title} {\bibinfo
  {title} {Device-independent randomness based on a tight upper bound of the
  maximal quantum value of chained inequality},\ }\href@noop {} {\bibfield
  {journal} {\bibinfo  {journal} {Physical Review A}\ }\textbf {\bibinfo
  {volume} {107}},\ \bibinfo {pages} {052415} (\bibinfo {year}
  {2023})}\BibitemShut {NoStop}%
\bibitem [{\citenamefont {Li}\ \emph {et~al.}(2019)\citenamefont {Li},
  \citenamefont {Gao}, \citenamefont {Cao},\ and\ \citenamefont
  {Wen}}]{li2019critical}%
  \BibitemOpen
  \bibfield  {author} {\bibinfo {author} {\bibfnamefont {D.-D.}\ \bibnamefont
  {Li}}, \bibinfo {author} {\bibfnamefont {F.}~\bibnamefont {Gao}}, \bibinfo
  {author} {\bibfnamefont {Y.}~\bibnamefont {Cao}},\ and\ \bibinfo {author}
  {\bibfnamefont {Q.-Y.}\ \bibnamefont {Wen}},\ }\bibfield  {title} {\bibinfo
  {title} {The critical detection efficiency for closing the detection loophole
  of some modified {B}ell inequalities},\ }\href@noop {} {\bibfield  {journal}
  {\bibinfo  {journal} {Quantum Information Processing}\ }\textbf {\bibinfo
  {volume} {18}},\ \bibinfo {pages} {1} (\bibinfo {year} {2019})}\BibitemShut
  {NoStop}%
\bibitem [{\citenamefont {Wooltorton}\ \emph {et~al.}(2022)\citenamefont
  {Wooltorton}, \citenamefont {Brown},\ and\ \citenamefont
  {Colbeck}}]{wooltorton2022tight}%
  \BibitemOpen
  \bibfield  {author} {\bibinfo {author} {\bibfnamefont {L.}~\bibnamefont
  {Wooltorton}}, \bibinfo {author} {\bibfnamefont {P.}~\bibnamefont {Brown}},\
  and\ \bibinfo {author} {\bibfnamefont {R.}~\bibnamefont {Colbeck}},\
  }\bibfield  {title} {\bibinfo {title} {Tight analytic bound on the trade-off
  between device-independent randomness and nonlocality},\ }\href@noop {}
  {\bibfield  {journal} {\bibinfo  {journal} {Physical Review Letters}\
  }\textbf {\bibinfo {volume} {129}},\ \bibinfo {pages} {150403} (\bibinfo
  {year} {2022})}\BibitemShut {NoStop}%
\bibitem [{\citenamefont {Watrous}(2011)}]{Watrous11}%
  \BibitemOpen
  \bibfield  {author} {\bibinfo {author} {\bibfnamefont {J.}~\bibnamefont
  {Watrous}},\ }\bibfield  {title} {\bibinfo {title} {{CS 867/QIC 890
  Semidefinite Programming in Quantum Information}}} (\bibinfo {year} {2011}),\
  \bibinfo {note} {lecture notes}\BibitemShut {NoStop}%
\bibitem [{\citenamefont {Fawzi}()}]{cvxquad}%
  \BibitemOpen
  \bibfield  {author} {\bibinfo {author} {\bibfnamefont {H.}~\bibnamefont
  {Fawzi}},\ }\href@noop {} {\bibinfo {title} {Rational lower/upper bounds on
  log}},\ \bibinfo {howpublished}
  {\url{https://github.com/hfawzi/cvxquad/blob/master/doc/log_approx_bounds.pdf}},\
  \bibinfo {note} {accessed: 2022-01-28}\BibitemShut {NoStop}%
\bibitem [{\citenamefont {V{\'e}rtesi}\ and\ \citenamefont
  {Bene}(2010)}]{vertesi2010two}%
  \BibitemOpen
  \bibfield  {author} {\bibinfo {author} {\bibfnamefont {T.}~\bibnamefont
  {V{\'e}rtesi}}\ and\ \bibinfo {author} {\bibfnamefont {E.}~\bibnamefont
  {Bene}},\ }\bibfield  {title} {\bibinfo {title} {Two-qubit {B}ell inequality
  for which positive operator-valued measurements are relevant},\ }\href@noop
  {} {\bibfield  {journal} {\bibinfo  {journal} {Physical Review A}\ }\textbf
  {\bibinfo {volume} {82}},\ \bibinfo {pages} {062115} (\bibinfo {year}
  {2010})}\BibitemShut {NoStop}%
\bibitem [{\citenamefont {Marangon}\ \emph {et~al.}(2017)\citenamefont
  {Marangon}, \citenamefont {Vallone},\ and\ \citenamefont
  {Villoresi}}]{marangon2017source}%
  \BibitemOpen
  \bibfield  {author} {\bibinfo {author} {\bibfnamefont {D.~G.}\ \bibnamefont
  {Marangon}}, \bibinfo {author} {\bibfnamefont {G.}~\bibnamefont {Vallone}},\
  and\ \bibinfo {author} {\bibfnamefont {P.}~\bibnamefont {Villoresi}},\
  }\bibfield  {title} {\bibinfo {title} {Source-device-independent ultrafast
  quantum random number generation},\ }\href@noop {} {\bibfield  {journal}
  {\bibinfo  {journal} {Physical review letters}\ }\textbf {\bibinfo {volume}
  {118}},\ \bibinfo {pages} {060503} (\bibinfo {year} {2017})}\BibitemShut
  {NoStop}%
\bibitem [{\citenamefont {Avesani}\ \emph {et~al.}(2018)\citenamefont
  {Avesani}, \citenamefont {Marangon}, \citenamefont {Vallone},\ and\
  \citenamefont {Villoresi}}]{avesani2018source}%
  \BibitemOpen
  \bibfield  {author} {\bibinfo {author} {\bibfnamefont {M.}~\bibnamefont
  {Avesani}}, \bibinfo {author} {\bibfnamefont {D.~G.}\ \bibnamefont
  {Marangon}}, \bibinfo {author} {\bibfnamefont {G.}~\bibnamefont {Vallone}},\
  and\ \bibinfo {author} {\bibfnamefont {P.}~\bibnamefont {Villoresi}},\
  }\bibfield  {title} {\bibinfo {title} {Source-device-independent
  heterodyne-based quantum random number generator at 17 {G}bps},\ }\href@noop
  {} {\bibfield  {journal} {\bibinfo  {journal} {Nature communications}\
  }\textbf {\bibinfo {volume} {9}},\ \bibinfo {pages} {5365} (\bibinfo {year}
  {2018})}\BibitemShut {NoStop}%
\bibitem [{\citenamefont {Bru{\ss}}\ \emph {et~al.}(2004)\citenamefont
  {Bru{\ss}}, \citenamefont {D'Ariano}, \citenamefont {Lewenstein},
  \citenamefont {Macchiavello}, \citenamefont {Sen}, \citenamefont {Sen} \emph
  {et~al.}}]{bruss2004distributed}%
  \BibitemOpen
  \bibfield  {author} {\bibinfo {author} {\bibfnamefont {D.}~\bibnamefont
  {Bru{\ss}}}, \bibinfo {author} {\bibfnamefont {G.~M.}\ \bibnamefont
  {D'Ariano}}, \bibinfo {author} {\bibfnamefont {M.}~\bibnamefont
  {Lewenstein}}, \bibinfo {author} {\bibfnamefont {C.}~\bibnamefont
  {Macchiavello}}, \bibinfo {author} {\bibfnamefont {A.}~\bibnamefont {Sen}},
  \bibinfo {author} {\bibfnamefont {U.}~\bibnamefont {Sen}}, \emph {et~al.},\
  }\bibfield  {title} {\bibinfo {title} {Distributed quantum dense coding},\
  }\href@noop {} {\bibfield  {journal} {\bibinfo  {journal} {Physical Review
  Letters}\ }\textbf {\bibinfo {volume} {93}},\ \bibinfo {pages} {210501}
  (\bibinfo {year} {2004})}\BibitemShut {NoStop}%
\bibitem [{\citenamefont {Prabhu}\ \emph {et~al.}(2013)\citenamefont {Prabhu},
  \citenamefont {Pati}, \citenamefont {Sen}, \citenamefont {Sen} \emph
  {et~al.}}]{prabhu2013exclusion}%
  \BibitemOpen
  \bibfield  {author} {\bibinfo {author} {\bibfnamefont {R.}~\bibnamefont
  {Prabhu}}, \bibinfo {author} {\bibfnamefont {A.~K.}\ \bibnamefont {Pati}},
  \bibinfo {author} {\bibfnamefont {A.}~\bibnamefont {Sen}}, \bibinfo {author}
  {\bibfnamefont {U.}~\bibnamefont {Sen}}, \emph {et~al.},\ }\bibfield  {title}
  {\bibinfo {title} {Exclusion principle for quantum dense coding},\
  }\href@noop {} {\bibfield  {journal} {\bibinfo  {journal} {Physical Review
  A}\ }\textbf {\bibinfo {volume} {87}},\ \bibinfo {pages} {052319} (\bibinfo
  {year} {2013})}\BibitemShut {NoStop}%
\bibitem [{\citenamefont {Pironio}\ \emph {et~al.}(2009)\citenamefont
  {Pironio}, \citenamefont {Ac{\'\i}n}, \citenamefont {Brunner}, \citenamefont
  {Gisin}, \citenamefont {Massar},\ and\ \citenamefont
  {Scarani}}]{pironio2009device}%
  \BibitemOpen
  \bibfield  {author} {\bibinfo {author} {\bibfnamefont {S.}~\bibnamefont
  {Pironio}}, \bibinfo {author} {\bibfnamefont {A.}~\bibnamefont {Ac{\'\i}n}},
  \bibinfo {author} {\bibfnamefont {N.}~\bibnamefont {Brunner}}, \bibinfo
  {author} {\bibfnamefont {N.}~\bibnamefont {Gisin}}, \bibinfo {author}
  {\bibfnamefont {S.}~\bibnamefont {Massar}},\ and\ \bibinfo {author}
  {\bibfnamefont {V.}~\bibnamefont {Scarani}},\ }\bibfield  {title} {\bibinfo
  {title} {Device-independent quantum key distribution secure against
  collective attacks},\ }\href@noop {} {\bibfield  {journal} {\bibinfo
  {journal} {New Journal of Physics}\ }\textbf {\bibinfo {volume} {11}},\
  \bibinfo {pages} {045021} (\bibinfo {year} {2009})}\BibitemShut {NoStop}%
\bibitem [{\citenamefont {Seguinard}\ \emph {et~al.}(2023)\citenamefont
  {Seguinard}, \citenamefont {Piveteau}, \citenamefont {Mironowicz},\ and\
  \citenamefont {Bourennane}}]{seguinard2023experimental}%
  \BibitemOpen
  \bibfield  {author} {\bibinfo {author} {\bibfnamefont {A.~J.-M.}\
  \bibnamefont {Seguinard}}, \bibinfo {author} {\bibfnamefont {A.}~\bibnamefont
  {Piveteau}}, \bibinfo {author} {\bibfnamefont {P.}~\bibnamefont
  {Mironowicz}},\ and\ \bibinfo {author} {\bibfnamefont {M.}~\bibnamefont
  {Bourennane}},\ }\bibfield  {title} {\bibinfo {title} {Experimental
  certification of more than one bit of quantum randomness in the two inputs
  and two outputs scenario},\ }\href@noop {} {\bibfield  {journal} {\bibinfo
  {journal} {New Journal of Physics}\ }\textbf {\bibinfo {volume} {25}},\
  \bibinfo {pages} {113022} (\bibinfo {year} {2023})}\BibitemShut {NoStop}%
\bibitem [{\citenamefont {Yang}\ \emph {et~al.}(2019)\citenamefont {Yang},
  \citenamefont {Horodecki},\ and\ \citenamefont
  {Winter}}]{yang2019distributed}%
  \BibitemOpen
  \bibfield  {author} {\bibinfo {author} {\bibfnamefont {D.}~\bibnamefont
  {Yang}}, \bibinfo {author} {\bibfnamefont {K.}~\bibnamefont {Horodecki}},\
  and\ \bibinfo {author} {\bibfnamefont {A.}~\bibnamefont {Winter}},\
  }\bibfield  {title} {\bibinfo {title} {Distributed private randomness
  distillation},\ }\href@noop {} {\bibfield  {journal} {\bibinfo  {journal}
  {Physical review letters}\ }\textbf {\bibinfo {volume} {123}},\ \bibinfo
  {pages} {170501} (\bibinfo {year} {2019})}\BibitemShut {NoStop}%
\bibitem [{\citenamefont {Cerf}\ and\ \citenamefont
  {Adami}(1999)}]{cerf1999quantum}%
  \BibitemOpen
  \bibfield  {author} {\bibinfo {author} {\bibfnamefont {N.~J.}\ \bibnamefont
  {Cerf}}\ and\ \bibinfo {author} {\bibfnamefont {C.}~\bibnamefont {Adami}},\
  }\bibfield  {title} {\bibinfo {title} {Quantum extension of conditional
  probability},\ }\href@noop {} {\bibfield  {journal} {\bibinfo  {journal}
  {Physical Review A}\ }\textbf {\bibinfo {volume} {60}},\ \bibinfo {pages}
  {893} (\bibinfo {year} {1999})}\BibitemShut {NoStop}%
\bibitem [{\citenamefont {Friis}\ \emph {et~al.}(2019)\citenamefont {Friis},
  \citenamefont {Vitagliano}, \citenamefont {Malik},\ and\ \citenamefont
  {Huber}}]{friis2019entanglement}%
  \BibitemOpen
  \bibfield  {author} {\bibinfo {author} {\bibfnamefont {N.}~\bibnamefont
  {Friis}}, \bibinfo {author} {\bibfnamefont {G.}~\bibnamefont {Vitagliano}},
  \bibinfo {author} {\bibfnamefont {M.}~\bibnamefont {Malik}},\ and\ \bibinfo
  {author} {\bibfnamefont {M.}~\bibnamefont {Huber}},\ }\bibfield  {title}
  {\bibinfo {title} {Entanglement certification from theory to experiment},\
  }\href@noop {} {\bibfield  {journal} {\bibinfo  {journal} {Nature Reviews
  Physics}\ }\textbf {\bibinfo {volume} {1}},\ \bibinfo {pages} {72} (\bibinfo
  {year} {2019})}\BibitemShut {NoStop}%
\bibitem [{\citenamefont {Cerf}\ \emph {et~al.}(1997)\citenamefont {Cerf},
  \citenamefont {Adami},\ and\ \citenamefont {Gingrich}}]{cerf1997quantum}%
  \BibitemOpen
  \bibfield  {author} {\bibinfo {author} {\bibfnamefont {N.~J.}\ \bibnamefont
  {Cerf}}, \bibinfo {author} {\bibfnamefont {C.}~\bibnamefont {Adami}},\ and\
  \bibinfo {author} {\bibfnamefont {R.~M.}\ \bibnamefont {Gingrich}},\
  }\bibfield  {title} {\bibinfo {title} {Quantum conditional operator and a
  criterion for separability},\ }\href@noop {} {\bibfield  {journal} {\bibinfo
  {journal} {arXiv preprint quant-ph/9710001}\ } (\bibinfo {year}
  {1997})}\BibitemShut {NoStop}%
\bibitem [{\citenamefont {Horodecki}\ \emph {et~al.}(1996)\citenamefont
  {Horodecki}, \citenamefont {Horodecki},\ and\ \citenamefont
  {Horodecki}}]{HORODECKI19961}%
  \BibitemOpen
  \bibfield  {author} {\bibinfo {author} {\bibfnamefont {M.}~\bibnamefont
  {Horodecki}}, \bibinfo {author} {\bibfnamefont {P.}~\bibnamefont
  {Horodecki}},\ and\ \bibinfo {author} {\bibfnamefont {R.}~\bibnamefont
  {Horodecki}},\ }\bibfield  {title} {\bibinfo {title} {Separability of mixed
  states: necessary and sufficient conditions},\ }\href
  {https://doi.org/https://doi.org/10.1016/S0375-9601(96)00706-2} {\bibfield
  {journal} {\bibinfo  {journal} {Physics Letters A}\ }\textbf {\bibinfo
  {volume} {223}},\ \bibinfo {pages} {1} (\bibinfo {year} {1996})}\BibitemShut
  {NoStop}%
\bibitem [{\citenamefont {Peres}(1996)}]{peres1996separability}%
  \BibitemOpen
  \bibfield  {author} {\bibinfo {author} {\bibfnamefont {A.}~\bibnamefont
  {Peres}},\ }\bibfield  {title} {\bibinfo {title} {Separability criterion for
  density matrices},\ }\href@noop {} {\bibfield  {journal} {\bibinfo  {journal}
  {Physical Review Letters}\ }\textbf {\bibinfo {volume} {77}},\ \bibinfo
  {pages} {1413} (\bibinfo {year} {1996})}\BibitemShut {NoStop}%
\bibitem [{pic()}]{picos}%
  \BibitemOpen
  \href@noop {} {\bibinfo {title} {A python interface to conic optimization
  solvers}},\ \bibinfo {howpublished}
  {\url{https://picos-api.gitlab.io/picos/index.html}},\ \bibinfo {note}
  {accessed: 2022-01-28}\BibitemShut {NoStop}%
\bibitem [{\citenamefont {Andersen}\ \emph {et~al.}(2011)\citenamefont
  {Andersen}, \citenamefont {Dahl}, \citenamefont {Liu}, \citenamefont
  {Vandenberghe}, \citenamefont {Sra}, \citenamefont {Nowozin},\ and\
  \citenamefont {Wright}}]{cvxopt}%
  \BibitemOpen
  \bibfield  {author} {\bibinfo {author} {\bibfnamefont {M.}~\bibnamefont
  {Andersen}}, \bibinfo {author} {\bibfnamefont {J.}~\bibnamefont {Dahl}},
  \bibinfo {author} {\bibfnamefont {Z.}~\bibnamefont {Liu}}, \bibinfo {author}
  {\bibfnamefont {L.}~\bibnamefont {Vandenberghe}}, \bibinfo {author}
  {\bibfnamefont {S.}~\bibnamefont {Sra}}, \bibinfo {author} {\bibfnamefont
  {S.}~\bibnamefont {Nowozin}},\ and\ \bibinfo {author} {\bibfnamefont {S.~J.}\
  \bibnamefont {Wright}},\ }\bibfield  {title} {\bibinfo {title}
  {Interior-point methods for large-scale cone programming},\ }\href@noop {}
  {\bibfield  {journal} {\bibinfo  {journal} {Optimization for machine
  learning}\ }\textbf {\bibinfo {volume} {5583}} (\bibinfo {year}
  {2011})}\BibitemShut {NoStop}%
\bibitem [{\citenamefont {ApS}(2019)}]{mosek}%
  \BibitemOpen
  \bibfield  {author} {\bibinfo {author} {\bibfnamefont {M.}~\bibnamefont
  {ApS}},\ }\href {http://docs.mosek.com/9.0/toolbox/index.html} {\emph
  {\bibinfo {title} {The {MOSEK} optimization toolbox for {MATLAB} manual.
  Version 9.0.}}} (\bibinfo {year} {2019})\BibitemShut {NoStop}%
\end{thebibliography}
\end{document}